\documentclass[journal, 12pt, onecolumn, draftclsnofoot]{IEEEtran}

\usepackage[utf8]{inputenc} 
\usepackage[T1]{fontenc}
\usepackage{url}
\usepackage{ifthen}
\usepackage{cite}
\usepackage[cmex10]{amsmath} 

\usepackage{color}
\usepackage{amssymb}
\usepackage{amsfonts}
\usepackage{tikz}
\usepackage{multirow}
\usepackage{multicol}
\usepackage{ctable}
\usepackage{mdframed}
\usepackage{colortbl}
\usepackage{pgfplots}
\usepackage{graphicx}
\usepackage{subcaption}
\usepackage{cancel}

\usepackage{verbatim}
\usepackage{stfloats}
\usepackage[bookmarks=false]{}
\usepackage{epsfig}
\usepackage{caption}
\usepackage{amssymb}
\usepackage{bm}
\usepackage{amsthm}
\usepackage{cite}
\usepackage{palatino}
\usepackage{arydshln}
\usetikzlibrary{arrows}
\usepackage{stfloats}

\usetikzlibrary{shapes,decorations,arrows,calc,arrows.meta,fit,positioning}

\newtheorem{theorem}{Theorem}
\newtheorem{lemma}{Lemma}

\DeclareCaptionLabelFormat{adja-page}{\hrulefill\\#1 #2 \emph{(previous page)}}

\usetikzlibrary{decorations.markings}
\colorlet{Mygreen}{black!60!green}
\colorlet{Myorange}{orange!40!gray}

\makeatletter
\newcommand{\thickhline}{%
	\noalign {\ifnum 0=`}\fi \hrule height 1.5pt
	\futurelet \reserved@a \@xhline
}
\newcolumntype{"}{@{\hskip\tabcolsep\vrule width 1pt\hskip\tabcolsep}}
\makeatother

\begin{document}
\title{GCSA Codes with Noise Alignment for Secure Coded Multi-Party Batch Matrix Multiplication}

\author{%
  \IEEEauthorblockN{Zhen Chen, Zhuqing Jia, Zhiying Wang and Syed A. Jafar}
  \IEEEauthorblockA{\\Center for Pervasive Communications and Computing (CPCC), UC Irvine\\
                    Email: \{zhenc4, zhuqingj, zhiying, syed\}@uci.edu}
}

\maketitle

\begin{abstract}
A secure multi-party batch matrix multiplication problem (SMBMM) is considered, where the goal is to allow a master to efficiently compute the pairwise products of two batches of massive matrices, by distributing the computation across S servers. Any X colluding servers gain no information about the input, and the master gains no additional information about the input beyond the product. A solution called Generalized Cross Subspace Alignment codes with Noise Alignment (GCSA- NA) is proposed in this work, based on cross-subspace alignment codes. The state of art solution to SMBMM is a coding scheme called polynomial sharing (PS) that was proposed by Nodehi and Maddah-Ali. GCSA-NA outperforms PS codes in several key aspects — more efficient and secure inter-server communication, lower latency, flexible inter-server network topology, efficient batch processing, and tolerance to stragglers. The idea of noise alignment can also be combined with N-source Cross Subspace Alignment (N-CSA) codes and fast matrix multiplication algorithms like Strassen's construction. Moreover, noise alignment can be applied to symmetric secure private information retrieval to achieve the asymptotic capacity.
\end{abstract}


\section{Introduction}
Recent interest in coding for secure, private, and distributed computing combines a variety of elements such as coded distributed massive matrix multiplication, straggler tolerance, batch computing and private information retrieval \cite{Lee_Lam_Pedarsani, Yu_Maddah-Ali_Avestimehr_Polynomial,Dutta_Fahim_Haddadpour, GPolyDot, Yu_Maddah-Ali_Avestimehr, Yu_Lagrange,Yu_cubic, Reisizadeh_Prakash_Pedarsani, Lee_Suh_Ramchandran, Dutta_Cadambe_Short, Dutta_Cadambe_Codedconv, Yu_Maddah-Ali_CodedDFT, Jahani-Nezhad_Maddah-Ali, Baharav_Lee_Ocal, Suh_Lee_Msparse, Wang_Liu_CLT, Mallick_Chaudhari_Joshi, Wang_Liu_Sparse, Severinson_iAmat_Rosnes, Haddadpour_Cadambe_Finite,Sheth_Dutta_Chaudhari, Jeong_Ye_Grover,Kim_Sohn_Moon_Group,Park_Lee_Sohn,Li_Maddah-Ali_Fog, Chang_Tandon, Kakar_Ebadifar_Sezgin_CSA, Oliveira_Rouayheb_Karpuk, Kim_Lee, Aliasgari_Simeone_Kliewer, Sun_Jafar_PIR,Sun_Jafar_TPIR, Banawan_Ulukus,Banawan_Ulukus_BPIR,Kadhe_Garcia_Heidarzadeh_Rouayheb_Sprintson,Wang_Skoglund_SPIRAd,Jia_Sun_Jafar_XSTPIR,Jia_Jafar_GXSTPIR,Jia_Jafar_MDSXSTPIR,Jia_Jafar_CDBC}. 
These related ideas converged recently in Generalized Cross Subspace Alignment (GCSA) codes presented in \cite{Jia_Jafar_CDBC}. GCSA codes originated in the setting of secure private information retrieval \cite{Jia_Sun_Jafar_XSTPIR} and  have recently been developed further in \cite{Jia_Jafar_CDBC} for applications to coded distributed batch computation problems. GCSA codes generalize and improve upon the state of art distributed computing schemes such as Polynomials codes \cite{Yu_Maddah-Ali_Avestimehr_Polynomial}, MatDot codes and PolyDot codes \cite{Dutta_Fahim_Haddadpour}, Generalized PolyDot codes \cite{GPolyDot} and Entangled Polynomial Codes \cite{Yu_Maddah-Ali_Avestimehr} that partition matrices into submatrices, as well as Lagrange Coded Computing \cite{Yu_Lagrange,Yu_cubic}  that allows batch processing of multiple computations.

As the next step in the expanding scope of coding for distributed computing, recently in \cite{Nodehi_Maddah_MPC} Nodehi and Maddah-Ali explored its application to secure multiparty computation \cite{Yao_first}. Specifically, Nodehi et al. consider a system including $N$ sources, $S$ servers and one master. Each source sends a coded function of its data (called a share) to each server. The servers process their inputs and while doing so, may communicate with each other. After that each server sends a message to the master, such that the master can recover the required function of the source inputs. The input data must be kept perfectly secure from the servers even if up to $X$ of the servers collude among themselves. The master must not gain any information about the input data beyond the result. 
Nodehi et al. propose a scheme called polynomial sharing (PS), which admits basic matrix operations such as addition and multiplication. By concatenating basic operations, arbitrary polynomial function can be calculated. The PS scheme has a few key limitations. It needs multiple rounds of communication among servers where every server needs to send messages to every other server. This carries a high communication cost and requires the network topology among servers to be a complete graph (otherwise data security may be compromised), does not tolerate stragglers, and does not lend itself to batch processing. These aspects (batch processing, improved inter-server communication efficiency, various network topologies) are highlighted as open problems by Nodehi et al. in \cite{Nodehi_Maddah_MPC}. 

Since GCSA codes are particularly efficient at batch processing and already encompass  prior approaches to coded distributed computing, in this work we explore whether GCSA codes can also be applied to the problem identified by Nodehi et al. In particular, we  focus on the problem of multiplication of two matrices. As it turns out, in this context the answer is in the affirmative. Securing the data against any $X$ colluding servers is already possible with GCSA codes as shown in \cite{Jia_Jafar_CDBC}. The only remaining challenge is how to prevent the master from learning anything about the inputs besides the result of the computation. Let us refer to the additional terms that are contained in the answers sent by the servers to the master, which may collectively reveal information about the inputs beyond the result of the computation, as \emph{interference} terms. To secure these interference terms, we use the idea of Noise Alignment (NA) -- the workers communicate among themselves to share  noise terms (unknown to the master) that are structured in the same manner as the interfering terms. Because of their matching structures, when added to the answer, the noise terms align perfectly with the interference terms and as a result no information is leaked to the master about the input data besides the result of the computation. Notably, the idea of noise alignment is not novel. While there are superficial distinctions, noise alignment is used essentially in the same manner in \cite{IA_NA}.

The combination of GCSA codes with noise alignment, GCSA-NA in short,  leads to significant advantages over PS schemes. Foremost, because it uses GCSA codes, it allows the benefits of batch processing as well as straggler robustness, neither of which are available in the PS scheme of  \cite{Nodehi_Maddah_MPC}. The only reason any inter-server communication is needed in a GCSA-NA scheme is to share the aligned noise terms among the servers. Since these terms do not depend on the data inputs, the inter-server communication in a GCSA-NA scheme is secure in a stronger sense than possible with PS, i.e., even if all inter-server communication is leaked, it can reveal nothing about the data inputs. In fact, the inter-server communication can take place \emph{before} the input data is determined, say during off-peak hours. This directly leads to another advantage. The GCSA-NA scheme allows the inter-server communication network  graph to be any connected graph unlike PS schemes which require a complete graph. 

The rest of the paper is organized as follows. Section \ref{s:ps} presents the problem statement. In Section \ref{s:res} we state the main result and compare it with previous approaches. A toy example is presented in Section \ref{toy}. The construction and proof of GCSA-NA are shown in Section \ref{s:con}. Section \ref{s:dis} concludes the paper.

\emph{Notation:} For positive integers $M, N$ ($M<N$), $[N]$ stands for the set $\{1,2,\dots,N\}$ and $[M:N]$ stands for the set $\{M,M+1,\dots,N\}$. For a set $\mathcal{I}=\{i_1,i_2,\dots,i_N\}$, $X_{\mathcal{I}}$ denotes the set $\{X_{i_1},X_{i_2},\dots,X_{i_N}\}$. The notation $\otimes$ denotes the Kronecker product of two matrices. $\mathbf{I}_N$ denotes the $N\times N$ identity matrix. $\mathbf{T}(X_1,X_2,\cdots,X_N)$ denotes the $N\times N$ lower triangular Toeplitz matrix, i.e.,
\small
\begin{align}
    \mathbf{T}(X_1,X_2,\cdots,X_N)=\begin{bmatrix}X_1&&&\\X_2&X_1& \\\vdots &\ddots &\ddots \\X_N&\cdots &X_2&X_1\end{bmatrix}. \notag
\end{align}
\normalsize
For a matrix $M$, $|M|$ denotes the number of elements in $M$. For a polynomial $P$, $deg_{\alpha}(P)$ denotes the degree with respect to a variable $\alpha$. Define the degree of the zero polynomial as $-1$. The notation $\widetilde{\mathcal{O}}(a\log^2b)$ suppresses polylog terms. It may be replaced with $\mathcal{O}(a\log^2b)$ if the field $\mathbb{F}$ supports the Fast Fourier Transform (FFT), and with $\mathcal{O}(a\log^2b\log\log(b))$ if it does not.

\section{Problem Statement} \label{s:ps}
\begin{figure}
\centering
   \includegraphics[width=.6\linewidth]{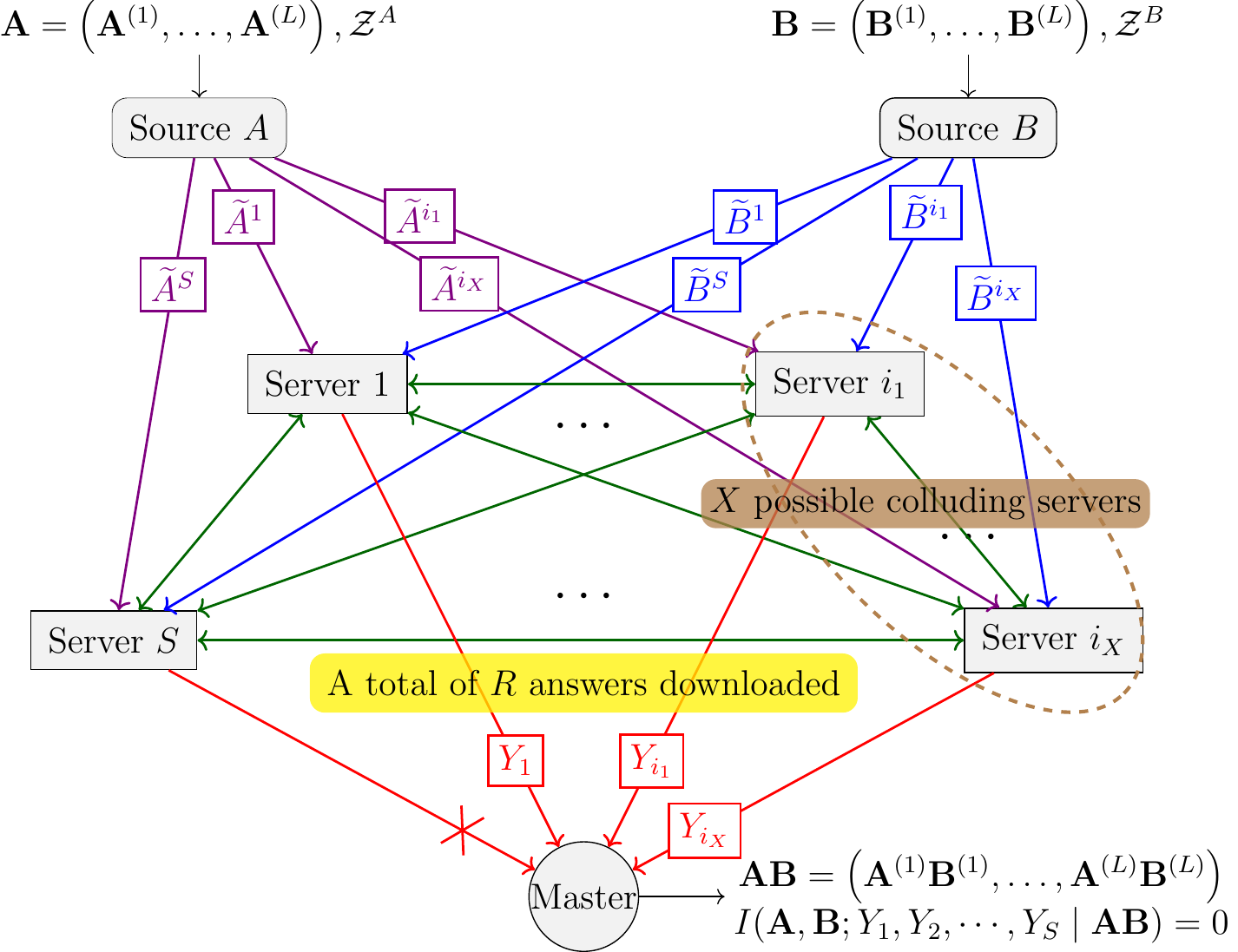}
\caption{\it The SMBMM problem. Sources generate matrices $\mathbf{A}=(\mathbf{A}^{(1)},\mathbf{A}^{(2)},\cdots,\mathbf{A}^{(L)})$ with separate noise $\mathcal{Z}^A$ and $\mathbf{B}=(\mathbf{B}^{(1)},\mathbf{B}^{(2)},\cdots,\mathbf{B}^{(L)})$ with separate noise $\mathcal{Z}^B$, and upload information to $S$ distributed servers in coded form $\widetilde{A}^{[S]}$, $\widetilde{B}^{[S]}$, respectively. Servers could send to each other some messages. For security, any $X$ colluding servers (e.g., Servers $i_1$ to $i_X$ in the figure) gain nothing about $\mathbf{A}, \mathbf{B}$. The $s^{th}$ server computes the answer $Y_s$, which is a function of all information available to it. For effective straggler (e.g., Server $S$ in the figure) mitigation, upon downloading answers from any $R$ servers, where $R<S$, the master must be able to recover the product $\mathbf{A}\mathbf{B}=(\mathbf{A}^{(1)}\mathbf{B}^{(1)},\mathbf{A}^{(2)}\mathbf{B}^{(2)},\dots,\mathbf{A}^{(L)}\mathbf{B}^{(L)})$. For privacy, the master must not gain any additional information about $\mathbf{A}, \mathbf{B}$ beyond the desired product $\mathbf{AB}$.}
    \label{fig:SMPC}
\end{figure}

Consider a system including $2$ sources ($A$ and $B$), $S$ servers (workers) and one master, as illustrated in Fig. \ref{fig:SMPC}. Each source is connected to every single server. Servers are connected to each other, and all of the servers are connected to the master. All of these links are secure and error free. 

Source $A$ and $B$ independently generate  sequences\footnote{The batch size $L$ can be chosen to be arbitrarily large by the coding algorithm.} of $L$ matrices, denoted as $\mathbf{A}=\left(\mathbf{A}^{(1)}, \mathbf{A}^{(2)},\dots, \mathbf{A}^{(L)}\right)$, and $\mathbf{B}=\left(\mathbf{B}^{(1)}, \mathbf{B}^{(2)},\dots, \mathbf{B}^{(L)}\right)$, respectively, such that $\forall l\in[L]$, $\mathbf{A}^{(l)}\in \mathbb{F}^{\lambda\times\kappa}$ and $\mathbf{B}^{(l)}\in\mathbb{F}^{\kappa\times\mu}$. The master is interested in the sequence of product matrices, $\mathbf{A}\mathbf{B}=\left(\mathbf{A}^{(1)}\mathbf{B}^{(1)}, \mathbf{A}^{(2)}\mathbf{B}^{(2)}, \dots,\mathbf{A}^{(L)}\mathbf{B}^{(L)}\right)$. The system operates in three phases: $1)$ sharing, $2)$ computation and communication, and $3)$ reconstruction. 

\emph{1) Sharing:} Each source encodes (encrypts) its matrices for the $s^{th}$ server as $\widetilde{A}^s$ and $\widetilde{B}^s$, so $
\widetilde{A}^s=f_s(\mathbf{A}, \mathcal{Z}^A), \widetilde{B}^s=g_s(\mathbf{B}, \mathcal{Z}^B)$, where $\mathcal{Z}^A$ and $\mathcal{Z}^B$ represent private randomness (noise) generated by the source. The encoded matrices, $\widetilde{A}^s,  \widetilde{B}^s$,  are sent to the $s^{th}$ server. 
	
\emph{2) Computation and Communication:} Servers may send messages to other servers, and process what they received from both the sources and other servers. Denote the communication from Server $s$ to Server $s'$ as $M_{s\rightarrow s'}$. Define $\mathcal{M}_s \triangleq \{M_{s'\rightarrow s} \mid s' \in [S]\setminus\{s\}\}$ as the messages that Server $s$ receives from other servers,  and $\mathcal{M} \triangleq \{\mathcal{M}_s \mid s \in [S]\}$  as the total messages that all servers receive. After the communication among servers, each server $s$ computes a response $Y_s$ and sends it to the master. $Y_s$ is a function of $\widetilde{A}^s$, $\widetilde{B}^s$ and $\mathcal{M}_s$, i.e., $Y_s=h_s(\widetilde{A}^s, \widetilde{B}^s, \mathcal{M}_s)$, where $h_s, s\in[S]$ are the functions used to produce the answer, and we denote them collectively as $\textit{\textbf{h}}=(h_1,h_2,\dots,h_S)$.
	
\emph{3) Reconstruction:} The master downloads information from servers. Some servers may fail to respond (or respond after the master executes the reconstruction), such servers are called stragglers. The master decodes the sequence of product matrices $\mathbf{A}\mathbf{B}$ based on the information from the responsive servers, using a class of decoding functions $\textit{\textbf{d}} =\{d_{\mathcal{R}} \mid \mathcal{R}\subset[S]\}$ where $d_{\mathcal{R}}$ is the decoding function used when the set of responsive servers is $\mathcal{R}$. 

This scheme must satisfy three constraints.

\textbf{Correctness:} The master must be able to recover the desired products $\mathbf{AB}$, i.e.,
	\begin{equation}
	H(\mathbf{AB} \mid Y_{\mathcal{R}})=0,
	\end{equation}
	or equivalently $\mathbf{AB}=d_{\mathcal{R}}(Y_{\mathcal{R}})$, for some $\mathcal{R}$.
	 
\textbf{Security \& Strong Security:} We first define \emph{security} which is called privacy for workers in  \cite{Nodehi_Maddah_MPC}. The servers must remain oblivious to the content of the data $\mathbf{A}, \mathbf{B}$, even if $X$ of them collude. Formally, $\forall \mathcal{X} \subset [S], |\mathcal{X}| \le X$,
	\begin{equation}
	I(\mathbf{A},  \mathbf{B} ; \widetilde{A}^{\mathcal{X}}, \widetilde{B}^{\mathcal{X}}, \mathcal{M}_{\mathcal{X}})=0,
	\end{equation}
	
In this paper, \emph{strong security} is also considered. It requires that the information transmitted among servers is independent of data $\mathbf{A}, \mathbf{B}$ and all the shares $\widetilde{A}^{\mathcal{[S]}}, \widetilde{B}^{\mathcal{[S]}}$, i.e., 
	\begin{equation}
	I(\mathbf{A},  \mathbf{B}, \widetilde{A}^{\mathcal{[S]}}, \widetilde{B}^{\mathcal{[S]}} ; \mathcal{M})=0.
	\end{equation}
	This property makes it possible that inter-server communications happen before receiving data from sources, and makes the server communication network topology more flexible. Note that PS does not satisfy strong security because $H\left( \mathbf{AB} \mid \mathcal{M} \right) = 0$ in the PS scheme.
	
\textbf{Privacy:} The master must not gain any additional information about $\mathbf{A}, \mathbf{B}$, beyond the required product. Precisely, 
	\begin{equation}
	I(\mathbf{A},  \mathbf{B} ;  Y_1, Y_2, \cdots, Y_S \mid \mathbf{AB})=0.
	\end{equation}
	This is the privacy for the master in \cite{Nodehi_Maddah_MPC}.
	
We say that $(\textit{\textbf{f}},\textit{\textbf{g}},\textit{\textbf{h}},\textit{\textbf{d}})$ form an SMBMM (\textbf{S}ecure coded \textbf{M}ulti-party \textbf{B}atch \textbf{M}atrix \textbf{M}ultiplication) code if it satisfies these three constraints. An SMBMM code is said to be $r$-recoverable if the master is able to recover the desired products from the answers obtained from any $r$ servers. In particular, an SMBMM code $(\textit{\textbf{f}},\textit{\textbf{g}},\textit{\textbf{h}},\textit{\textbf{d}})$ is $r$-recoverable if for any $\mathcal{R}\subset[S]$, $|\mathcal{R}|=r$, and for any realization of $\mathbf{A}$, $\mathbf{B}$, we have $\mathbf{AB}=d_{\mathcal{R}}(Y_{\mathcal{R}})$. Define the recovery threshold $R$ of an SMBMM code $(\textit{\textbf{f}},\textit{\textbf{g}},\textit{\textbf{h}},\textit{\textbf{d}})$ to be the minimum integer $r$ such that the SMBMM code is $r$-recoverable.

The communication cost of an SMBMM code is comprised of these parts: upload cost of the sources, communication cost among the servers, and download cost of the master. The (normalized)\footnote{We normalize source upload cost with the number of elements contained in the constituent matrices ${\bf A}, {\bf B}$. The server communication cost and master download cost  are normalized by the number of elements contained in the desired product ${\bf AB}$.} upload costs $U_A$ and $U_B$ are defined as follows.
\begin{align}
    U_A=\frac{\sum_{s\in[S]}|\widetilde{A}^s|}{L\lambda\kappa}, ~~U_B=\frac{\sum_{s\in[S]}|\widetilde{B}^s|}{L\kappa\mu}.
\end{align}
Similarly, the (normalized) server communication cost $CC$ and download cost $D$ are defined as follows.
\begin{align}
	CC=\frac{|\mathcal{M}|}{L\lambda\mu}, ~~ D=\max_{\mathcal{R}, \mathcal{R}\subset[S], |\mathcal{R}|=R}\frac{\sum_{s\in\mathcal{R}}|Y_s|}{L\lambda\mu}.
\end{align}

Next let us consider the complexity of encoding, decoding and server computation. Define the (normalized) computational complexity at each server, $\mathcal{C}_s$, to be the order of the number of arithmetic operations required to compute the function $h_s$ at each server, normalized by $L$. Similarly, define the (normalized) encoding computational complexity $\mathcal{C}_{eA}$ for $\widetilde{A}^{[S]}$ and $\mathcal{C}_{eB}$ for $\widetilde{B}^{[S]}$ as the order of the number of arithmetic operations required to compute the functions $\textit{\textbf{f}}$ and $\textit{\textbf{g}}$, respectively, each normalized by $L$. Finally, define the (normalized) decoding computational complexity $\mathcal{C}_d$ to be the order of the number of arithmetic operations required to compute $d_{\mathcal{R}}(Y_{\mathcal{R}})$, maximized over $\mathcal{R}, \mathcal{R}\subset[S], |\mathcal{R}|=R$, and normalized by $L$. Note that normalization by batch-size $L$ is needed to have fair comparisons between batch processing approaches and individual matrix-partitioning solutions \emph{per matrix multiplication}.

\section{Main Result}  \label{s:res}
Our main result appears in the following theorem.
\begin{theorem}\label{thm:gcsa}
For SMBMM over a field $\mathbb{F}$ with $S$ servers, $X$-security, and positive integers $(\ell, K_c, p, m, n)$ such that $m\mid\lambda$, $p\mid\kappa$, $n\mid\mu$ and $L=\ell K_c\leq|\mathbb{F}|-S$, the GCSA-NA scheme presented in Section \ref{s:con} is a solution, and its recovery threshold, cost, and complexity are listed as follows.
\begin{align*}
    \text{Recovery Threshold:} && R&=pmn(\ell+1)K_c+2X-1,\\
    \text{Source Upload Cost of $\widetilde{A}^{[S]},\widetilde{B}^{[S]}$:}&& (U_A, U_B)&=\left(\frac{S}{K_cpm},\frac{S}{K_cpn}\right),\\
    \text{Server Communication Cost:}&&  CC &= \frac{S-1}{\ell K_cmn},\\
    \text{Master Download Cost:} && D&=\frac{R}{\ell K_cmn},\\
    \text{Source Encoding Complexity for $\widetilde{A}^{[S]},\widetilde{B}^{[S]}$:}&& (\mathcal{C}_{eA},\mathcal{C}_{eB}) &=\left(\widetilde{\mathcal{O}}\left(\frac{\lambda\kappa S\log^2S}{K_cpm}\right),  \widetilde{\mathcal{O}}\left(\frac{\kappa\mu S\log^2S}{K_cpn}\right)\right),\\
    \text{Server Computation Complexity:}&& \mathcal{C}_s &=\mathcal{O}\left(\frac{\lambda\kappa\mu}{K_cpmn}\right),\\
    \text{Master Decoding Complexity:}&& \mathcal{C}_d&=\widetilde{\mathcal{O}}\left(\lambda\mu p\log^2R\right).
\end{align*}
\end{theorem}

\noindent The following observations place the result of Theorem \ref{thm:gcsa} in perspective.

1. GCSA-NA codes are based on the construction of GCSA codes from \cite{Jia_Jafar_CDBC}, combined with the idea of noise-alignment (e.g., \cite{IA_NA}). In turn, GCSA codes are based on a combination of CSA codes for batch processing \cite{Jia_Jafar_CDBC} and EP codes for matrix partitioning \cite{Yu_Maddah-Ali_Avestimehr}. CSA codes are themselves based on the idea of cross-subspace alignment (CSA) that was introduced in the context of secure PIR \cite{Jia_Sun_Jafar_XSTPIR}. It is a remarkable coincidence that while the idea of CSA originated in the context of PIR \cite{Jia_Sun_Jafar_XSTPIR}, and Lagrange Coded Computing was introduced in parallel independently in \cite{Yu_Lagrange} for the context of coded computing, the two approaches are essentially identical, with CSA codes being slightly more powerful in the context of coded distributed matrix multiplication (CSA codes offer additional improvements over LCC codes in terms of download cost \cite{Jia_Jafar_CDBC}). Indeed, LCC codes for batch matrix multiplication are recovered as a special case of CSA codes.

2. The idea of noise alignment can be applied to the $N$-CSA codes \cite{Jia_Jafar_CDBC}, for $N$-source secure coded multi-party batch matrix computation. In \cite{Yu_cubic}, Strassen’s construction \cite{Strassen}, combined with LCC, are introduced for batch distributed matrix multiplication. Noise alignment is also applicable to Strassen's constructions (see Section \ref{s:dis}). By setting $K_c=1$, $\ell=L$ and $S=R$, the construction of GCSA-NA codes, with a straightforward generalization, can be further modified to settle the asymptotic (the number of message goes to infinity) capacity of symmetric $X$-secure $T$-private computation (and also the corresponding private information retrieval setting) \cite{Jia_Sun_Jafar_XSTPIR}. However, the amount of randomness required by the construction is not necessarily optimal. For example, it is shown in \cite{Jia_Sun_Jafar_XSTPIR} that by the achievable scheme for XSTPIR,  symmetric security (privacy) is automatically satisfied when $T=1$, i.e., no randomness among servers is required. 

\begin{table*}[!htbp]
    \centering
    \begin{tabular}{rcc}\specialrule{.2em}{.1em}{.1em}\rowcolor{white}
    &   Polynomial Sharing (PS \cite{Nodehi_Maddah_MPC}) &  GCSA-NA \\\specialrule{.2em}{.1em}{.1em}\rowcolor{white}
    Strong Security & No & Yes\\ \hline
     Recovery Threshold $(R)$ &$2pmn+2X-1$ & $pmn(\ell+1)K_c+2X-1$\\\hline

      Straggler Tolerance & No $(S=R)$ & Yes. Tolerates $S-R$ stragglers\\ \hline
        Server Network Topology & Complete Graph & Any Connected Graph\\\hline
     \begin{tabular}{rr}Source Encoding\\ Complexity \end{tabular}$(\mathcal{C}_{eA}, \mathcal{C}_{eB})$ &$\left(\widetilde{\mathcal{O}}\left(\frac{\lambda\kappa S\log^2S}{pm}\right),  \widetilde{\mathcal{O}}\left(\frac{\kappa\mu S\log^2S}{pn}\right)\right)$ & $\left(\widetilde{\mathcal{O}}\left(\frac{\lambda\kappa S\log^2S}{K_cpm}\right),  \widetilde{\mathcal{O}}\left(\frac{\kappa\mu S\log^2S}{K_cpn}\right)\right)$ \\\hline 
   Source Upload Cost $(U_A, U_B)$&$\left(\frac{S}{pm}, \frac{S}{pn}\right)$ &$\left(\frac{S}{K_cpm}, \frac{S}{K_cpn}\right)$\\\hline

     \begin{tabular}{rr}Server Communication Cost \end{tabular} $(CC)$ &$\frac{S(S-1)}{mn}$ & $\frac{S-1}{\ell K_cmn}$ \\\hline
   \begin{tabular}{rr}Server Computation\\ Complexity \end{tabular} $(\mathcal{C}_{s})$  & $\mathcal{O}\left(\frac{\lambda\kappa\mu}{pmn}\right)+\mathcal{O}\left(\lambda\mu\right)+\widetilde{\mathcal{O}}\left(\frac{S\log^2 S\lambda\mu}{mn}\right)$ & $\mathcal{O}\left(\frac{\lambda\kappa\mu}{K_cpmn}\right)+\mathcal{O}\left(\frac{\lambda\mu}{K_cmn}\right)+\widetilde{\mathcal{O}}\left(\frac{\lambda\mu\log^2 S}{\ell K_c mn}\right)$\\
    & $+\mathcal{O}\left(\frac{(S-1)\lambda\mu}{mn}\right)\approx \mathcal{O}\left(\frac{\lambda\kappa\mu}{pmn}\right)$ if $\frac{\kappa}{p}\gg S$ & $\approx\mathcal{O}\left(\frac{\lambda\kappa\mu}{K_cpmn}\right)$  if $\frac{\kappa}{p}\gg S$ \\\hline
     \begin{tabular}{rr}Master Download Cost \end{tabular}$(D)$ &$\frac{mn+X}{mn}$ &$\frac{R}{\ell K_cmn}$ \\\hline
  \begin{tabular}{rr}Master Decoding Complexity \end{tabular}$(\mathcal{C}_d)$&$\widetilde{\mathcal{O}}\left(\lambda\mu \log^2(mn+X)\right)$&$\widetilde{\mathcal{O}}\left(\lambda\mu p \log^2(R)\right)$\\
  \specialrule{.2em}{.1em}{.1em}
    \end{tabular}
      \caption{\it \small Performance Comparison of PS and \emph{GCSA-NA}.}
    \label{tab:full}

 \end{table*}
 
\begin{figure}
\centering
   \begin{subfigure}{0.49\linewidth}
   \centering
   \includegraphics[width=\linewidth]{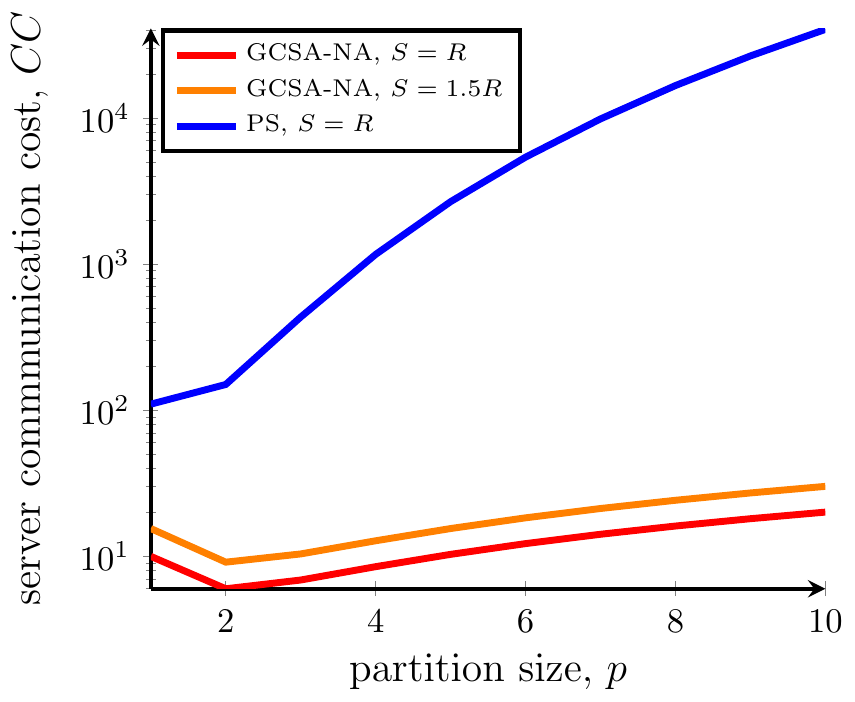}
   \caption{}
   \label{fig:cca} 
\end{subfigure}
\hfill
\begin{subfigure}{0.49\linewidth}
   \centering
   \includegraphics[width=\linewidth]{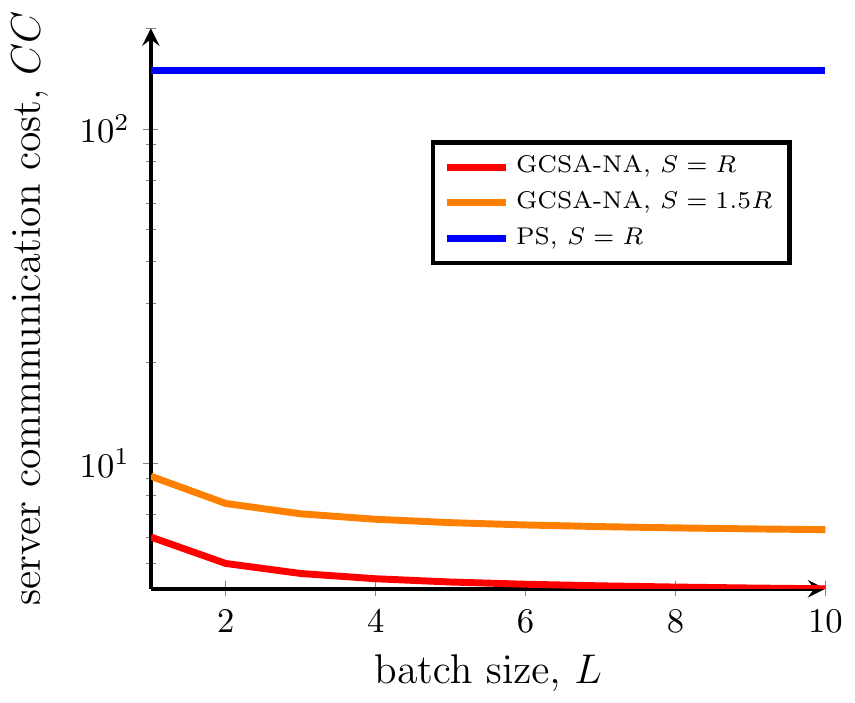}
   \caption{}
   \label{fig:ccb}
\end{subfigure}
\centering
\caption{$\lambda = \kappa = \mu$, $p=m=n$. (a) Server communication cost vs. partition size, given $L=1$ and $X=5$. (b) Server communication cost vs. batch size, given $p=2$ and $X=5$. }
\end{figure}

3. A side-by-side comparison of the GCSA-NA solution with polynomial sharing (PS) appears in Table \ref{tab:full}. Because all inter-server communication is independent of input data, GCSA-NA schemes are strongly secure, i.e., even if all inter-server communication is leaked it does not compromise the security of input data. In GCSA-NA the inter-server network graph can be any connected graph. This is not possible with PS. For example, if the inter-server network graph is a star graph, then the hub server can decode $\mathbf{AB}$ by monitoring all the inter-server communication in a PS scheme, violating the security constraint. Unlike the PS scheme, in GCSA-NA, all inter-server communication can take place during off-peak hours, even before the input data is generated, giving GCSA-NA a significant latency advantage. Unlike PS where every server must communicate with every server, i.e., $S(S-1)$ such inter-server communications must take place, GCSA-NA only requires $S-1$ inter-server communications to propagate structured noise terms across all servers. This improvement is shown numerically in Fig. \ref{fig:cca}. The server computation complexity is also lower for the GCSA-NA scheme than the PS scheme. This is because in PS, each server needs to multiply the two shares received from the sources, calculate the shares for \emph{every other server} and sum up all the shares from \emph{every other server}. However, in GCSA-NA, each server only needs to multiply the two shares received from the sources and add noise (which can be precomputed during off-peak hours). This advantage is particularly significant for large number of servers. The GCSA-NA scheme naturally allows robustness to stragglers, which is particularly important for massive matrix multiplications. Stragglers can be an especially significant concern for PS because of the strongly sequential nature of multi-round computation that is central to PS. This is because server failures between computation rounds disrupt the computation sequence. Remarkably, Fig. \ref{fig:cca} shows that the inter-server communication cost of GCSA-NA is significantly better than PS even when GCSA-NA accommodates stragglers (while PS does not).

When restricted to batch size $1$, i.e., with $\ell=K_c=1$, GCSA-NA has the same recovery threshold as PS. Now consider batch processing, i.e., batch size $L>1$, e.g., with $L=K_c, \ell=1$. PS can be applied to batch processing by repeating the scheme $L$ times. Fig. \ref{fig:ccb} shows that the normalized server communication cost of GCSA-NA decreases as $L$ increases and is significantly less than that in PS. For the same number of servers $S$, the upload cost of GCSA-NA is smaller by a factor of $1/K_c$ compared to PS. GCSA-NA does have higher download cost and decoding complexity than PS by approximately a factor of $p$, which depends on how the matrices are partitioned. If $p$ is a small value, e.g., $p=1$, then the costs are quite similar. The improvement in download cost and decoding complexity of PS by a factor of $1/p$ comes at the penalty of increased inter-server communication cost by a factor of $S$. But since $S\geq R\geq 2pmn+2X-1\geq p$, and typically $S\gg p$, the improvement is dominated by the penalty, so that overall the communication cost of PS is still significantly higher. 

\section{Toy Example} \label{toy}
Let us consider a toy example with parameters $\lambda = \kappa = \mu, m=n=1, p=2, l=1, K_c=2, X=1$ and $S=R$. Suppose matrices $\mathbf{A}, \mathbf{B} \in \mathbb{F}^{\lambda\times\lambda}$, and we wish to multiply matrix $\mathbf{A}=[\mathbf{A}_1 ~ \mathbf{A}_2]$ with matrix $\mathbf{B}=\left[\begin{matrix}\mathbf{B}_1^T ~ \mathbf{B}_2^T\end{matrix}\right]^T$ to compute the product $\mathbf{AB}=\mathbf{A}_1\mathbf{B}_1+\mathbf{A}_2\mathbf{B}_2$, where $\mathbf{A}_1, \mathbf{A}_2 \in \mathbb{F}^{\lambda\times\frac{\lambda}{2}}, \mathbf{B}_1, \mathbf{B}_2 \in \mathbb{F}^{\frac{\lambda}{2}\times\lambda}$. For this toy example we summarize both the Polynomial Sharing approach \cite{Nodehi_MPC_EP_ISIT, Nodehi_MPC_EP, Nodehi_Maddah_MPC}, and our GCSA-NA approach.
\subsection{Polynomial Sharing Solution} \label{pss}
Polynomial sharing is based on EP code \cite{Yu_Maddah-Ali_Avestimehr} . The given partitioning corresponds to EP code construction for $m=n=1, p=2$, and we have
\begin{align}
P&= \mathbf{A}_1+\alpha \mathbf{A}_2, \quad Q=\alpha \mathbf{B}_1+\mathbf{B}_2 \label{ep1}\\
\implies PQ&= \mathbf{A}_1\mathbf{B}_2+\alpha(\mathbf{A}_1\mathbf{B}_1+\mathbf{A}_2\mathbf{B}_2)+\alpha^2\mathbf{A}_2\mathbf{B}_1.
\end{align} 

To satisfy $X=1$ security, PS includes noise with each share, i.e., $\widetilde{A}=P+\alpha^2\mathbf{Z}^A, \widetilde{B}=Q+\alpha^2\mathbf{Z}^B,$ where  $\alpha, \widetilde{A}, \widetilde{B}$ are generic variables that should be replaced with $\alpha_s,\widetilde{A}^s, \widetilde{B}^s$ for Server $s$, and $\alpha_1, \cdots, \alpha_S$ are distinct elements. Each server computes the product of the shares that it receives, i.e., 
\small
\begin{align}
&\qquad \widetilde{A}\widetilde{B}=PQ + \alpha^2P\mathbf{Z}^B + \alpha^2\mathbf{Z}^AQ + \alpha^4\mathbf{Z}^A\mathbf{Z}^B\\
&=\mathbf{A}_1\mathbf{B}_2+\alpha(\mathbf{A}_1\mathbf{B}_1+\mathbf{A}_2\mathbf{B}_2)+\alpha^2(\mathbf{A}_2\mathbf{B}_1+\mathbf{A}_1\mathbf{Z}^B + \mathbf{Z}^A\mathbf{B}_2) +\alpha^3(\mathbf{A}_2\mathbf{Z}^B + \mathbf{Z}^A\mathbf{B}_1) + \alpha^4\mathbf{Z}^A\mathbf{Z}^B.
\end{align}
\normalsize

To secure inputs from the master, PS requires that every server sends to the master only the desired term $\mathbf{A}_1\mathbf{B}_1+\mathbf{A}_2\mathbf{B}_2$ by using secret sharing scheme among servers. Since $deg_{\alpha}(\widetilde{A}\widetilde{B})=4$,  $\mathbf{A}_1\mathbf{B}_1+\mathbf{A}_2\mathbf{B}_2$ can be calculated from $5$ distinct $\widetilde{A}\widetilde{B}$ according to the Lagrange interpolation rules. In particular, there exist $5$ constants $r_1,\cdots,r_5$, such that $\mathbf{A}_1\mathbf{B}_1+\mathbf{A}_2\mathbf{B}_2 = \sum_{s\in[5]} r_s\widetilde{A}^s\widetilde{B}^s.$ Consider Server $s$, it sends $M_{s \rightarrow  j} = r_s\widetilde{A}^s\widetilde{B}^s + \alpha_j\mathbf{Z}_s$ to Server $j$, where $\mathbf{Z}_1, \cdots, \mathbf{Z}_5$ are i.i.d. uniform noise matrices. After Server $s$ collects all the shares $M_{j \rightarrow s}$, it sums them up 
\begin{align}
Y_s &= \sum_{j \in [5]} M_{j \rightarrow s} = \sum_{j \in [5]}r_j\widetilde{A}^j\widetilde{B}^j + \alpha_s\sum_{j \in [5]}\mathbf{Z}_j  = \mathbf{A}_1\mathbf{B}_1+\mathbf{A}_2\mathbf{B}_2 + \alpha_s\sum_{j \in [5]}\mathbf{Z}_j,
\end{align}
and sends $Y_s$ to the master. Note that after receiving $M_{j \rightarrow s}$ for all $j\in[5]$, Server s still gains no information about the input data, which guarantees the security. However, it does not satisfy strong security, because $\mathbf{AB}$ can be decoded based on $M_{j \rightarrow s}, j,s \in [5]$.

The master can decode $\mathbf{AB}$ after collecting $2$ responses from servers.\footnote{In \cite{Nodehi_MPC_EP}, for arbitrary polynomials, $M_{s \rightarrow  j} = r_s\widetilde{A}^s\widetilde{B}^s + \alpha_j^2\mathbf{Z}_s$ because $Y_s$ is forced to be casted in the form of entangled polynomial sharing.} Note that PS needs at least $S=R=5$ servers, since $5$ distinct $\widetilde{A}\widetilde{B}$ are required to obtain $Y_s$. 

\subsection{GCSA-NA Solution}
GCSA codes \cite{Jia_Jafar_CDBC} can handle batch processing, therefore let us consider batch size $2$ $(\ell=1, K_c=2)$. Denote the second instance by $\mathbf{A}',\mathbf{B}'$. Using CSA code,
\begin{align}
P&=\mathbf{A}_1+(f-\alpha)\mathbf{A}_2,\quad Q=(f-\alpha)\mathbf{B}_1+\mathbf{B}_2.\\
P'&=\mathbf{A}_1'+(f'-\alpha)\mathbf{A}_2', \quad Q'=(f'-\alpha)\mathbf{B}_1'+\mathbf{B}_2'.
\end{align}
and the shares are constructed as follows,
\begin{align}
\widetilde{A}=\Delta\left(\frac{P}{(f-\alpha)^2}+\frac{P'}{(f'-\alpha)^2}\right), \quad \widetilde{B}=\frac{Q}{(f-\alpha)^2}+\frac{Q'}{(f'-\alpha)^2}
\end{align}
where $\Delta = (f-\alpha)^2(f'-\alpha)^2$, and $\alpha, \widetilde{A}, \widetilde{B}$ are generic variables that should be replaced with $\alpha_s, \widetilde{A}^s, \widetilde{B}^s$ for Server $s$. Furthermore, $f, f', \alpha_1, \alpha_2, \cdots, \alpha_S$ are distinct elements.  Each server computes the product of the shares that it receives, i.e., 
\begin{align}
&\widetilde{A}\widetilde{B}=\frac{c_0}{(f-\alpha)^2}PQ+\frac{c_1}{f-\alpha}PQ+\frac{c_0'}{(f'-\alpha)^2}P'Q' +\frac{c_1'}{f'-\alpha}P'Q'+I_0+\alpha I_1+\alpha^2 I_2\\
&=\frac{c_0\mathbf{A}_1\mathbf{B}_2}{(f-\alpha)^2}+\frac{c_0\mathbf{A}_1\mathbf{B}_1+c_0\mathbf{A}_2\mathbf{B}_2+c_1\mathbf{A}_1\mathbf{B}_2}{f-\alpha} +\frac{c_0' \mathbf{A}_1'\mathbf{B}_2'}{(f'-\alpha)^2} +\frac{c_0'\mathbf{A}_1'\mathbf{B}_1'+c_0'\mathbf{A}_2'\mathbf{B}_2'+c_1'\mathbf{A}_1'\mathbf{B}_2'}{f'-\alpha}  \notag \\
&\qquad +I_0+\alpha I_1+\alpha^2 I_2,
\end{align}
where $I_0,I_1, I_2$ are combinations of $PQ, P'Q', PQ', P'Q$ and $c_0,c_1,c'_0,c'_1$ are constants. This is the original GCSA code \cite{Jia_Jafar_CDBC}, and we need $R=pmn((\ell+1)K_c-1)+p-1=7$ responses to recover the desired product.  

Next, let us modify the scheme to make it $X=1$ secure by including noise with each share, i.e.,
\begin{align}
\widetilde{A}&=\Delta\left(\frac{P}{(f-\alpha)^2}+\frac{P'}{(f'-\alpha)^2}+\mathbf{Z}^A\right), \quad \widetilde{B}=\frac{Q}{(f-\alpha)^2}+\frac{Q'}{(f'-\alpha)^2}+\mathbf{Z}^B. \\
\implies & \quad \widetilde{A}\widetilde{B}=\frac{c_0PQ}{(f-\alpha)^2}+\frac{c_1PQ}{f-\alpha}+\frac{c_0'P'Q'}{(f'-\alpha)^2}+\frac{c_1'P'Q'}{f'-\alpha} + \sum_{i=0}^4 \alpha^i I_i.
\end{align}
As a result of the added noise terms, the recovery threshold is now increased to $9$. Note that the term $I_4$ contains only contributions from $\Delta \mathbf{Z}^A \mathbf{Z}^B$, i.e., this term leaks no information about $\mathbf{A},\mathbf{B}$ matrices.

If the servers directly return their computed values of $\widetilde{A}\widetilde{B}$ to the master, then besides the result of the computation some additional information about the input matrices $\mathbf{A},\mathbf{B}$ may be leaked by the interference terms
\begin{align}
&\left(\frac{c_0}{(f-\alpha)^2}+\frac{c_1}{f-\alpha}\right)\mathbf{A}_1\mathbf{B}_2+\left(\frac{c_0'}{(f'-\alpha)^2}+\frac{c_1'}{f'-\alpha}\right)\mathbf{A}_1'\mathbf{B}_2' + \sum_{i=0}^3 \alpha^i I_i
\end{align}
which can be secured by the addition of \emph{aligned noise} terms
\begin{align}
\widetilde{Z}&=\left(\frac{c_0}{(f-\alpha)^2}+\frac{c_1}{f-\alpha}\right)\mathbf{Z}+\left(\frac{c_0'}{(f'-\alpha)^2}+\frac{c_1'}{f'-\alpha}\right)\mathbf{Z}'  + \sum_{i=0}^3 \alpha^i \mathbf{Z}_i
\end{align}
at each server so that the answer returned by each server to the master is $\widetilde{A}\widetilde{B}+\widetilde{Z}$. Here $\mathbf{Z}, \mathbf{Z}', \mathbf{Z}_0, \mathbf{Z}_1, \mathbf{Z}_2, \mathbf{Z}_3$ are i.i.d. uniform noise matrices, that can all be privately generated by one server, who can then share their aligned form $\widetilde{Z}$ with all other servers. This sharing of $\widetilde{Z}$ is the only inter-server communication needed in GCSA-NA. Since it is independent of the inputs, it can be done during off-peak hours, thereby reducing the latency of server computation. The strong security is also automatically satisfied.

\section{Construction of GCSA-NA} \label{s:con}
Now let us present the general construction. $L=\ell K_c$ instances of $\mathbf{A}$ and $\mathbf{B}$ matrices are split into $\ell$ groups. $\forall l\in[\ell], \forall k\in[K_c]$, denote
\begin{align}
    \mathbf{A}^{l,k}=\mathbf{A}^{(K_c(l-1)+k)},~ \mathbf{B}^{l,k}=\mathbf{B}^{(K_c(l-1)+k)}.
\end{align}
Further, each matrix $\mathbf{A}^{l,k}$ is partitioned into $m\times p$ blocks and each matrix $\mathbf{B}^{l,k}$ is partitioned into $p\times n$ blocks, i.e.,
\begin{align*}
\mathbf{A}^{l,k} = \left[\begin{array}{c c c c}
\mathbf{A}^{l,k}_{1,1} & \mathbf{A}^{l,k}_{1,2}  & \cdots & \mathbf{A}^{l,k}_{1,p}  \\
\mathbf{A}^{l,k}_{2,1} & \mathbf{A}^{l,k}_{2,2}  & \cdots & \mathbf{A}^{l,k}_{2,p}  \\
\vdots & \vdots & \vdots &  \vdots \\
\mathbf{A}^{l,k}_{m,1} & \mathbf{A}^{l,k}_{m,2}  & \cdots & \mathbf{A}^{l,k}_{m,p}  \\
\end{array}\right], 
\mathbf{B}^{l,k} = \left[\begin{array}{c c c c}
\mathbf{B}^{l,k}_{1,1} & \mathbf{B}^{l,k}_{1,2}  & \cdots & \mathbf{B}^{l,k}_{1,n}  \\
\mathbf{B}^{l,k}_{2,1} & \mathbf{B}^{l,k}_{2,2}  & \cdots & \mathbf{B}^{l,k}_{2,n}  \\
\vdots & \vdots & \vdots &  \vdots \\
\mathbf{B}^{l,k}_{p,1} & \mathbf{B}^{l,k}_{p,2}  & \cdots & \mathbf{B}^{l,k}_{p,n}  \\
\end{array}\right],
\end{align*}
where $\left(\mathbf{A}^{l,k}_{i,j}\right)_{i \in[m], j\in[p]} \in \mathbb{F}^{\frac{\lambda}{m}\times\frac{\kappa}{p}}$ and $\left(\mathbf{B}^{l,k}_{i,j}\right)_{i \in[m], j\in[p]} \in \mathbb{F}^{\frac{\kappa}{p}\times\frac{\mu}{n}}$.

Let $f_{1,1},f_{1,2},\cdots,f_{\ell,K_c},\alpha_1,\alpha_2,\cdots,\alpha_S$ be $(S+L)$ distinct elements from the field $\mathbb{F}$.  For convenience, define
\begin{align}
&R'=pmn, \quad D_E = \max(pm, pmn-pm+p)-1,\\
&\mathcal{E} = \left\{p+p(m'-1)+pm(n''-1) \mid m'\in[m], n''\in[n]\right\},\\
&\Delta_s^{l,K_c}=\prod_{k\in[K_c]}(f_{l,k}-\alpha_s)^{R'}, \forall l\in[\ell], \forall s\in[S].
\end{align}
Define $c_{l,k,i}, i\in\{0,1,\cdots,R'(K_c-1)\}$ to be the coefficients satisfying
\begin{align}
    \Psi_{l,k}(\alpha)&=\prod_{k'\in[K_c]\setminus\{k\}}\left(\alpha+(f_{l,k'}-f_{l,k})\right)^{R'} =\sum_{i=0}^{R'(K_c-1)}c_{l,k,i}\alpha^i, \forall l\in[\ell], \forall k\in[K_c], \label{cdef}
\end{align}
i.e., they are the coefficients of the polynomial $\Psi_{l,k}(\alpha)=\prod_{k'\in[K_c]\setminus\{k\}}\left(\alpha+(f_{l,k'}-f_{l,k})\right)^{R'}$, which is defined by its roots. Note that all the coefficients $(c_{l,k,i})_{l\in[L],k\in[K_c],i\in\{0,1,\cdots,R'(K_c-1)\}}$, $\alpha_{[S]}$, $(f_{l,k})_{l\in[L],k\in[K]}$ are globally known.

\subsection{Sharing}
Firstly, each source encodes each constituent matrix blocks $\mathbf{A}^{l,k}$ and $\mathbf{B}^{l,k}$ with Entangled Polynomial code  \cite{Yu_Maddah-Ali_Avestimehr}. For all $l\in[\ell],k\in[K_c]$, define
\begin{align}
    P_s^{l,k}&=\sum_{m'\in[m]}\sum_{p'\in[p]}\mathbf{A}^{l,k}_{m',p'}(f_{l,k}-\alpha_s)^{p'-1+p(m'-1)},\label{ept1}\\
    Q_s^{l,k}&=\sum_{p''\in[p]}\sum_{n''\in[n]}\mathbf{B}^{l,k}_{p'',n''}(f_{l,k}-\alpha_s)^{p-p''+pm(n''-1)}. \label{ept2}
\end{align}
Note that the original Entangled Polynomial code can be regarded as polynomials of $\alpha_s$, and here for each $(l,k)$, Entangled Polynomial code is constructed as polynomials of $(f_{l,k}-\alpha_s)$.

Each source generates $\ell X$ independent random matrices, $\mathcal{Z}^A=\left\{\mathbf{Z}_{1,1}^{A}, \cdots, \mathbf{Z}_{\ell,X}^{A}\right\}$ and $\mathcal{Z}^B=\left\{\mathbf{Z}_{1,1}^{B}, \cdots, \mathbf{Z}_{\ell,X}^{B}\right\}$. 
The independence is established as follows.
\begin{align}
H(\mathcal{Z}^A, \mathcal{Z}^B, \mathbf{A}, \mathbf{B}) &= H(\mathbf{A})+H( \mathbf{B}) + \sum_{l\in[\ell],x\in[X]}H\left(\mathbf{Z}_{l,x}^{A}\right) +\sum_{l\in[\ell],x\in[X]}H\left(\mathbf{Z}_{l,x}^{B}\right). \label{noise1}
\end{align}
For all $s\in[S]$, the shares of matrices $\mathbf{A}$ and $\mathbf{B}$ at the $s^{th}$ server are constructed as $\widetilde{A}^s=(\widetilde{A}_{1}^s,\widetilde{A}_{2}^s,\dots,\widetilde{A}_{\ell}^s), \widetilde{B}^s=(\widetilde{B}_{1}^s,\widetilde{B}_{2}^s,\dots,\widetilde{B}_{\ell}^s)$,
where for all $l\in[\ell]$,
\small
\begin{align}
    \widetilde{A}_{l}^s=\Delta_s^{l,K_c}\left(\sum_{k\in[K_c]}\frac{P_s^{l,k}}{(f_{l,k}-\alpha_s)^{R'}}+ \sum_{x\in[X]}\alpha_s^{x-1}\mathbf{Z}_{l,x}^{A}\right), \quad \widetilde{B}_{l}^s=\sum_{k\in[K_c]}\frac{Q_s^{l,k}}{(f_{l,k}-\alpha_s)^{R'}} +\sum_{x\in[X]}\alpha_s^{x-1}\mathbf{Z}_{l,x}^{B}.
\end{align}
\normalsize
Then each pair of shares $\widetilde{A}^s, \widetilde{B}^s$ is sent to the corresponding server. 

\subsection{Computation and Communication}
One of the servers generates a set of $\frac{\lambda}{m}\times\frac{\mu}{n}$ matrices $\mathcal{Z}^{server}$, which contains $R'(K_c-1)+X+D_E+\ell K_c(p-1)mn$ independent random matrices and $\ell K_cmn$ zero matrices. In particular, $\mathcal{Z}^{server} = \left\{\mathcal{Z}^{server}_1, \mathcal{Z}^{server}_2\right\}$, $\mathcal{Z}^{server}_1=\left\{\mathbf{Z}'_i \mid i\in[R'(K_c-1)+X+D_E]\right\}$, and $\mathcal{Z}^{server}_2=\left\{\mathbf{Z}''_{l,k,i} \mid l\in[\ell], k\in[K_c], i\in[R'] \right\}$. Here,
\begin{align*}
\mathbf{Z}''_{l,k,i} = 
\begin{cases}
\mathbf{0}, & \text{if $i\in\mathcal{E}$}\\
\mathbf{Z}'''_{l,k,i}, & \text{otherwise,}
\end{cases} ~\forall l\in[\ell], \forall k\in[K_c].
\end{align*}
Here $\mathbf{Z}'_i$ and $\mathbf{Z}'''_{l,k,i}$ are the independent random matrices. The independence is established as follows.
\small
\begin{align}
H(\mathcal{Z}^{server}, \mathbf{A}, \mathbf{B}) &= H(\mathbf{A})+H( \mathbf{B}) + \sum_{i\in[R'(K_c-1)+X+D_E]}H(\mathbf{Z}'_i) +\sum_{l\in[\ell], k\in[K_c], i\in[R']}H(\mathbf{Z}''_{l,k,i}). \label{noise2}
\end{align}
\normalsize
Without loss of generality, assume the first server generates $\mathcal{Z}^{server}$, encodes them into 
\begin{align}
\widetilde{M}_s &= \sum_{x\in[R'(K_c-1)+X+D_E]}\alpha_s^{x-1}\mathbf{Z}'_{x} + \sum_{l\in[\ell]}\sum_{k\in[K_c]}\sum_{i=0}^{R'-1}\frac{\sum_{i'=0}^{i}c_{l,k.i-i'}\mathbf{Z}''_{l,k,i'+1}}{(f_{l,k}-\alpha_s)^{R'-i}}, \label{cr}
\end{align}
and sends $\widetilde{M}_s$ to server $s, s\in[S]\setminus\{1\}$, where $c_{l,k,i}$ 
is defined in \eqref{cdef}. The answer returned by the $s^{th}$ server to the master is constructed as $Y_s=\sum_{l\in[\ell]}\widetilde{A}^s_{l}\widetilde{B}^s_{l} + \widetilde{M}_s$.

\subsection{Reconstruction}
After the master collects any $R$ answers, it decodes the desired products $\mathbf{AB}$. 

\subsection{Proof of Theorem 1}
To begin, let us recall the standard result for Confluent Cauchy-Vandermonde matrices \cite{Gasca_Martinez_Muhlbach}, replicated here for the sake of completeness.
\begin{lemma}\label{lemma:ccv}
If $f_{1,1}, f_{1,2}, \cdots, f_{\ell,K_c}, \alpha_1,\alpha_2,\cdots,\alpha_R$ are $R+L$ distinct elements of $\mathbb{F}$, with $|\mathbb{F}|\geq R+L$, $L=\ell K_c$ and $R=R'(\ell+1)K_c+2X-1$, then the $R\times R$ Confluent Cauchy-Vandermonde matrix \eqref{matrix} is invertible over $\mathbb{F}$.
\small
\begin{align}
\hat{{\bf V}}_{\ell,K_c,R',X,R}&\triangleq\left[
\begin{matrix}
\frac{1}{(f_{1,1}-\alpha_1)^{R'}}&\cdots&\frac{1}{f_{1,1}-\alpha_1}&\cdots&\frac{1}{(f_{\ell,K_c}-\alpha_1)^{R'}}&\cdots&\frac{1}{f_{\ell,K_c}-\alpha_1}&1&\cdots&\alpha_1^{R'K_c+2X-2}\\
\frac{1}{(f_{1,1}-\alpha_2)^{R'}}&\cdots&\frac{1}{f_{1,1}-\alpha_2}&\cdots&\frac{1}{(f_{\ell,K_c}-\alpha_2)^{R'}}&\cdots&\frac{1}{f_{\ell,K_c}-\alpha_2}&1&\cdots&\alpha_2^{R'K_c+2X-2}\\
\vdots&\vdots&\vdots&\vdots&\vdots&\vdots&\vdots&\vdots&\vdots&\vdots\\
\frac{1}{(f_{1,1}-\alpha_R)^{R'}}&\cdots&\frac{1}{f_{1,1}-\alpha_R}&\cdots&\frac{1}{(f_{\ell,K_c}-\alpha_R)^{R'}}&\cdots&\frac{1}{f_{\ell,K_c}-\alpha_R}&1&\cdots&\alpha_R^{R'K_c+2X-2}\\
\end{matrix}
\right] \label{matrix}
\end{align}
\normalsize
\end{lemma}

Firstly, let us prove that the GCSA-NA codes are $R=pmn(\ell+1)K_c+2X-1$ recoverable. Rewrite $Y_s$ as follows.
\small
\begin{align}
    &\quad Y_s=\widetilde{A}^s_{1}\widetilde{B}^s_{1}+\widetilde{A}^s_{2}\widetilde{B}^s_{2}+\dots+\widetilde{A}^s_{\ell}\widetilde{B}^s_{\ell} + \widetilde{M}_s \\
    &=\sum_{l\in[\ell]}\Delta_s^{l,K_c}\left(\sum_{k\in[K_c]}\frac{P_s^{l,k}}{(f_{l,k}-\alpha_s)^{R'}}+\sum_{x\in[X]}\alpha_s^{x-1}\mathbf{Z}_{l,x}^{A}\right)  \left(\sum_{k\in[K_c]}\frac{Q_s^{l,k}}{(f_{l,k}-\alpha_s)^{R'}}+\sum_{x\in[X]}\alpha_s^{x-1}\mathbf{Z}_{l,x}^{B}\right) + \widetilde{M}_s \\
    &=\sum_{l\in[\ell]}\Delta_s^{l,K_c}\left(\sum_{k\in[K_c]}\frac{P_s^{l,k}}{(f_{l,k}-\alpha_s)^{R'}}\right) \left(\sum_{k\in[K_c]}\frac{Q_s^{l,k}}{(f_{l,k}-\alpha_s)^{R'}}\right) + \underbrace{\sum_{l\in[\ell]}\Delta_s^{l,K_c}\left(\sum_{k\in[K_c]}\frac{P_s^{l,k}}{(f_{l,k}-\alpha_s)^{R'}}\right) \left(\sum_{x\in[X]}\alpha_s^{x-1}\mathbf{Z}_{l,x}^{B}\right)}_{\Gamma_2} \notag \\
    &\quad + \underbrace{\sum_{l\in[\ell]}\Delta_s^{l,K_c}\left(\sum_{k\in[K_c]}\frac{Q_s^{l,k}}{(f_{l,k}-\alpha_s)^{R'}}\right) \left(\sum_{x\in[X]}\alpha_s^{x-1}\mathbf{Z}_{l,x}^{A}\right)}_{\Gamma_3} + \underbrace{\sum_{l\in[\ell]}\Delta_s^{l,K_c} \left(\sum_{x\in[X]}\alpha_s^{x-1}\mathbf{Z}_{l,x}^{A}\right) \left(\sum_{x\in[X]}\alpha_s^{x-1}\mathbf{Z}_{l,x}^{B}\right)}_{\Gamma_4}+ \widetilde{M}_s  \label{eq:anscorrv0} \\  
    &=\sum_{l\in[\ell]}\sum_{k\in[K_c]}\frac{\prod_{k'\in[K_c]\setminus\{k\}}(f_{l,k'}-\alpha_s)^{R'}}{(f_{l,k}-\alpha_s)^{R'}}P_s^{l,k}Q_s^{l,k}+\underbrace{\sum_{l\in[\ell]}\sum_{\substack{k,k'\in[K_c]\\k\neq k'}}\left(\prod_{k''\in[K_c]\setminus\{k,k'\}}(f_{l,k''}-\alpha_s)^{R'}\right)P_s^{l,k}Q_s^{l,k'}}_{\Gamma_1} \notag \\
    &\quad + \Gamma_2 + \Gamma_3 + \Gamma_4 + \widetilde{M}_s.\label{eq:anscorrv}
\end{align}
\normalsize
Consider the first term in \eqref{eq:anscorrv}. For each $l\in[\ell], k\in[K_c]$, we have
\small
\begin{align}
    &\qquad \frac{\prod_{k'\in[K_c]\setminus\{k\}}(f_{l,k'}-\alpha_s)^{R'}}{(f_{l,k}-\alpha_s)^{R'}}P_s^{l,k}Q_s^{l,k} \notag \\
    &=\frac{\prod_{k'\in[K_c]\setminus\{k\}}\left((f_{l,k}-\alpha_s)+(f_{l,k'}-f_{l,k})\right)^{R'}}{(f_{l,k}-\alpha_s)^{R'}}P_s^{l,k}Q_s^{l,k}\\
    &=\frac{\Psi_{l,k}(f_{l,k}-\alpha_s)}{(f_{l,k}-\alpha_s)^{R'}}P_s^{l,k}Q_s^{l,k}\label{eq:vcsagalpha}\\
    &=\left(\frac{c_{l,k,0}}{(f_{l,k}-\alpha_s)^{R'}}+\frac{c_{l,k,1}}{(f_{l,k}-\alpha_s)^{R'-1}}+\cdots+\frac{c_{l,k,R'-1}}{f_{l,k}-\alpha_s}\right)P_s^{l,k}Q_s^{l,k}\notag\\
    &\quad+\underbrace{\left(\sum_{i=R'}^{R'(K_c-1)}c_{l,k,i}(f_{l,k}-\alpha_s)^{i-R'}\right)P_s^{l,k}Q_s^{l,k}}_{\Gamma_5}.\label{eq:anscorrv1}
\end{align}
\normalsize
where \eqref{eq:vcsagalpha} results from the definition of $\Psi_{l,k}(\cdot)$ as in \eqref{cdef} and in \eqref{eq:anscorrv1} the polynomial $\Psi_{l,k}(f_{l,k}-\alpha_s)$ is rewritten in terms of its coefficients.  

By the construction of Entangled Polynomial code \eqref{ept1} \eqref{ept2}, the product $P_s^{l,k}Q_s^{l,k}$ can be written as weighted sums of the terms $1,(f_{l,k}-\alpha_s),\cdots,(f_{l,k}-\alpha_s)^{R'+p-2}$, i.e.,
\begin{align}
    P_s^{l,k}Q_s^{l,k} = \sum_{i=0}^{R'+p-2}\mathbf{C}^{l,k}_{i+1}(f_{l,k}-\alpha_s)^i, \label{eq:pq}
\end{align}
where $\mathbf{C}^{l,k}_{1},\mathbf{C}^{l,k}_{2},\cdots,\mathbf{C}^{l,k}_{R'+p-1}$ are various linear combinations of products of blocks of $\mathbf{A}^{l,k}$ and blocks of $\mathbf{B}^{l,k}$. 
Consider the first term in \eqref{eq:anscorrv1}.
\begin{align}
    &\left(\frac{c_{l,k,0}}{(f_{l,k}-\alpha_s)^{R'}}+\frac{c_{l,k,1}}{(f_{l,k}-\alpha_s)^{R'-1}}+\cdots+\frac{c_{l,k,R'-1}}{f_{l,k}-\alpha_s}\right)P_s^{l,k}Q_s^{l,k} \notag \\
    \stackrel{\eqref{eq:pq}}{=}&\left(\frac{c_{l,k,0}}{(f_{l,k}-\alpha_s)^{R'}}+\frac{c_{l,k,1}}{(f_{l,k}-\alpha_s)^{R'-1}}+\cdots+\frac{c_{l,k,R'-1}}{f_{l,k}-\alpha_s}\right)  \sum_{i=0}^{R'+p-2}\mathbf{C}^{l,k}_{i+1}(f_{l,k}-\alpha_s)^i\\
    =&\sum_{i=0}^{R'-1}\frac{\sum_{i'=0}^{i}c_{l,k,i-i'}\mathbf{C}^{l,k}_{i'+1}}{(f_{l,k}-\alpha_s)^{R'-i}} + \underbrace{\sum_{i=0}^{p-2}(f_{l,k}-\alpha_s)^i\left(\sum_{i'=i+1}^{R'+i'}c_{l,k,R'-i'+i}\mathbf{C}^{l,k}_{i'+1}\right)}_{\Gamma_6}  \notag \\
    &\quad + \underbrace{\sum_{i=p-1}^{R'+p-3}(f_{l,k}-\alpha_s)^i\left(\sum_{i'=i+1}^{R'+p-2}c_{l,k,R'-i'+i}\mathbf{C}^{l,k}_{i'+1}\right)}_{\Gamma_7}. \label{eq:anscorrv2}
\end{align}
Note that if $K_c=1$, $\forall i \neq 0, c_{l,k,i}=0$, then $\Gamma_5$ and $\Gamma_7$ are zero polynomials. Now let us consider the degree with respect to $\alpha_s$ of $\Gamma_1, \cdots, \Gamma_7$. 
\small
\begin{align*}
deg_{\alpha_s}\left( \Gamma_1 \right) &=
\begin{cases}
R'(K_c-1)+p-2, & \text{if $K_c>1$}\\
-1, & \text{otherwise}
\end{cases},  && deg_{\alpha_s}\left( \Gamma_2 \right) = R'(K_c-1)+pm+X-2,\\
deg_{\alpha_s}\left( \Gamma_3 \right) &= R'(K_c-1)+pmn-pm+p+X-2, && deg_{\alpha_s}\left( \Gamma_4 \right) = R'K_c+2X-2, ~~ deg_{\alpha_s}\left( \Gamma_6 \right) = p-2, \\\\
deg_{\alpha_s}\left( \Gamma_5 \right) &=\begin{cases}
R'(K_c-1)+p-2, & \text{if $K_c>1$}\\
-1, & \text{otherwise}
\end{cases},  && deg_{\alpha_s}\left( \Gamma_7 \right) =
\begin{cases}
R'+p-3, & \text{if $K_c>1$}\\
-1, & \text{otherwise}
\end{cases}.
\end{align*}
\normalsize

Recall $X, p, m, n, K_c$ are positive integers. If $K_c>1$, it is easy to see that $R'K_c+2X-2$ is the largest. If $K_c=1$, $R'=pmn\ge p>p-2$, $R'K_c+2X-2$ is also the largest. Therefore the sum of $\Gamma_1, \cdots, \Gamma_7$ can be expanded into weighted sums of the terms $1, \alpha_s, \cdots, \alpha_s^{R'K_c+2X-2}$. Note that the weights of terms $\alpha_s^{R'(K_c-1)+X+D_E+1}, \cdots, \alpha_s^{R'K_c+2X-2}$  are functions of $\mathcal{Z}^{A}, \mathcal{Z}^{B}$. $Y_s$ can be rewritten as
\begin{align}
&\qquad Y_s = \sum_{l\in[\ell]}\sum_{k\in[K_c]}\sum_{i=0}^{R'-1}\frac{\sum_{i'=0}^{i}c_{l,k,i-i'}\mathbf{C}^{l,k}_{i'+1}}{(f_{l,k}-\alpha_s)^{R'-i}} + \sum_{x\in[R'K_c+2X-1]}\alpha_s^{x-1}I_x +\widetilde{M}_s \\
	&\stackrel{\eqref{cr}}{=}\sum_{l\in[\ell]}\sum_{k\in[K_c]}\sum_{i=0}^{R'-1}\frac{\sum_{i'=0}^{i}c_{l,k,i-i'}\mathbf{C}^{l,k}_{i'+1}}{(f_{l,k}-\alpha_s)^{R'-i}}  + \sum_{x\in[R'K_c+2X-1]}\alpha_s^{x-1}I_x  +\sum_{x\in[R'(K_c-1)+X+D_E]}\alpha_s^{x-1}\mathbf{Z}'_{x} \notag \\
	& \quad + \sum_{l\in[\ell]}\sum_{k\in[K_c]}\sum_{i=0}^{R'-1}\frac{\sum_{i'=0}^{i}c_{l,k.i-i'}\mathbf{Z}''_{l,k,i'+1}}{(f_{l,k}-\alpha_s)^{R'-i}} \\
	&=\sum_{l\in[\ell]}\sum_{k\in[K_c]}\sum_{i=0}^{R'-1}\frac{\sum_{i'=0}^{i}c_{l,k,i-i'}\left(\mathbf{C}^{l,k}_{i'+1}+\mathbf{Z}''_{l,k,i'+1}\right)}{(f_{l,k}-\alpha_s)^{R'-i}} + \sum_{x\in[R'(K_c-1)+X+D_E]}\alpha_s^{x-1}\left(I_x +\mathbf{Z}'_{x}\right) \notag \\
	&\quad +  \sum_{x=R'(K_c-1)+X+D_E+1}^{R'K_c+2X-1}\alpha_s^{x-1}I_x \\
	&=\sum_{l\in[\ell]}\sum_{k\in[K_c]}\sum_{i=0}^{R'-1}\frac{\sum_{i'=0}^{i}c_{l,k,i-i'}\mathbf{D}^{l,k}_{i'+1}}{(f_{l,k}-\alpha_s)^{R'-i}} + \sum_{x\in[R'K_c+2X-1]}\alpha_s^{x-1}J_x,
\end{align}
where $\mathbf{D}^{l,k}_{i} = \mathbf{C}^{l,k}_{i}+\mathbf{Z}''_{l,k,i}, l\in[\ell], k\in[K_c], i\in[R']$,  $J_x = I_x +\mathbf{Z}'_{x}, x \in[R'(K_c-1)+X+D_E]$ and $J_x = I_x, x\in [R'(K_c-1)+X+D_E+1:R'K_c+2X-1]$. In the matrix form, answers from any $R=R'K_c+2X-1+R' L=pmn(\ell+1)K_c+2X-1$ servers, whose indices are denoted as $s_1,s_2,\cdots,s_R$, can be written as \eqref{mt}.
\small
\begin{align}
    &\begin{bmatrix}
    Y_{s_1}\\
    Y_{s_2}\\
    \vdots\\
    Y_{s_R}
    \end{bmatrix}= \underbrace{\left[
\begin{array}{ccc;{4pt/4pt}c;{4pt/4pt}ccc;{4pt/4pt}ccc}
\frac{1}{(f_{1,1}-\alpha_{s_1})^{R'}}&\cdots&\frac{1}{f_{1,1}-\alpha_{s_1}}&\cdots&\frac{1}{(f_{\ell,K_c}-\alpha_{s_1})^{R'}}&\cdots&\frac{1}{f_{\ell,K_c}-\alpha_{s_1}}&1&\cdots&\alpha_{s_1}^{R'K_c+2X-2}\\
\frac{1}{(f_{1,1}-\alpha_{s_2})^{R'}}&\cdots&\frac{1}{f_{1,1}-\alpha_{s_2}}&\cdots&\frac{1}{(f_{\ell,K_c}-\alpha_{s_2})^{R'}}&\cdots&\frac{1}{f_{\ell,K_c}-\alpha_{s_2}}&1&\cdots&\alpha_{s_2}^{R'K_c+2X-2}\\
\vdots&\vdots&\vdots&\vdots&\vdots&\vdots&\vdots&\vdots&\vdots&\vdots\\
\frac{1}{(f_{1,1}-\alpha_{s_R})^{R'}}&\cdots&\frac{1}{f_{1,1}-\alpha_{s_R}}&\cdots&\frac{1}{(f_{\ell,K_c}-\alpha_{s_R})^{R'}}&\cdots&\frac{1}{f_{\ell,K_c}-\alpha_{s_R}}&1&\cdots&\alpha_{s_R}^{R'K_c+2X-2}\\
\end{array}
\right]}_{\hat{\mathbf{V}}_{\ell,K_c,R',X, R}} \notag\\
    &\underbrace{\left[\begin{array}{c;{4pt/4pt}c;{4pt/4pt}c;{4pt/4pt}c}
    \mathbf{T}(c_{1,1,0},\cdots,c_{1,1,R'-1})&&&\\\hdashline[4pt/4pt]
    &\ddots&&\\\hdashline[4pt/4pt]
    &&\mathbf{T}(c_{\ell,K_c,0},\cdots,c_{\ell,K_c,R'-1})&\\\hdashline[4pt/4pt]
    &&&\mathbf{I}_{R-R' L}
    \end{array}\right]}_{\hat{\mathbf{V}}'_{\ell,K_c,R',X, R}} \otimes\mathbf{I}_{\lambda/m} \left[\begin{array}{c}
    \mathbf{D}^{1,1}_{1}\\
    \vdots\\
    \mathbf{D}^{1,1}_{R'}\\\hdashline[4pt/4pt]
    \vdots\\\hdashline[4pt/4pt]
    \mathbf{D}^{\ell,K_c}_{1}\\
    \vdots\\
    \mathbf{D}^{\ell,K_c}_{R'}\\\hdashline[4pt/4pt]
    J_1\\
    \vdots\\
    J_{R'K_c+2X-1}
    \end{array}\right]. \label{mt}
\end{align}
\normalsize
Since $f_{1,1}, f_{1,2}, \cdots, f_{\ell,K_c}$ are distinct, for all $l\in[\ell],k\in[K_c], c_{l,k,0}=\prod_{k'\in[K_c]\setminus\{k\}}(f_{l,k'}-f_{l,k})^{R'}$ are non-zero. Hence, the lower triangular toeplitz matrices $\mathbf{T}(c_{1,1,0},\cdots,c_{1,1,R'-1}),$ $\cdots,$ $\mathbf{T}(c_{\ell,K_c,0},\cdots,c_{\ell,K_c,R'-1})$ are non-singular, and the block diagonal matrix $\hat{\mathbf{V}}'_{\ell,K_c,R',X,R}$ is invertible. Guaranteed by Lemma \ref{lemma:ccv} and the fact that the Kronecker product of non-singular matrices is non-singular, the matrix $(\hat{\mathbf{V}}_{\ell,K_c,R',X,R}\hat{\mathbf{V}}'_{\ell,K_c,R',X,R})\otimes\mathbf{I}_{\lambda/m}$ is invertible. Therefore, the master is able to recover $\left(\mathbf{D}^{l,k}_{i}\right)_{l\in[\ell],k\in[K_c],i\in[R']}$ by inverting the matrix. Note that $\mathbf{Z}''_{l,k,i} = \mathbf{0}, l\in[\ell],k\in[K_c],i\in\mathcal{E}$, therefore $\left(\mathbf{C}^{l,k}_{i}\right)_{l\in[\ell],k\in[K_c],i\in\mathcal{E}}=\left(\mathbf{D}^{l,k}_{i}\right)_{l\in[\ell],k\in[K_c],i\in\mathcal{E}}$. The desired products $(\mathbf{A}^{(l)}\mathbf{B}^{(l)})_{l\in[L]}$ are recoverable from $\left(\mathbf{C}^{l,k}_{i}\right)_{l\in[\ell],k\in[K_c],i\in\mathcal{E}}$, guaranteed by the correctness of Entangled Polynomial code \cite{Yu_Maddah-Ali_Avestimehr}. This completes the proof of recovery threshold $R=pmn(\ell+1)K_c+2X-1$. 

Consider the strong security property. According to the construction, $\mathcal{M}_1=0$, $\mathcal{M}_s = \widetilde{M}_s, s\in[S]\setminus\{1\}$, and $\mathcal{M} = \{\widetilde{M}_s \mid s\in[S]\setminus\{1\}\}$. Since $\widetilde{M}_s$ is a function of $\mathcal{Z}^{server}$,
\begin{align}
I(\mathbf{A},  \mathbf{B}, \widetilde{A}^{\mathcal{[S]}}, \widetilde{B}^{\mathcal{[S]}}  ; \mathcal{M}) \le I(\mathbf{A},  \mathbf{B}, \widetilde{A}^{\mathcal{[S]}}, \widetilde{B}^{\mathcal{[S]}}  ; \mathcal{Z}^{server}) = 0.
\end{align}
Strong security is satisfied. Security is guaranteed because $\forall \mathcal{X} \subset [S], |\mathcal{X}|=X$,
\begin{align}
 I(\mathbf{A},  \mathbf{B} ; \widetilde{A}^{\mathcal{X}}, \widetilde{B}^{\mathcal{X}}, \mathcal{M}_{\mathcal{X}}) &= I(\mathbf{A},  \mathbf{B} ;\mathcal{M}_{\mathcal{X}}) + I(\mathbf{A},  \mathbf{B} ; \widetilde{A}^{\mathcal{X}}, \widetilde{B}^{\mathcal{X}} \mid \mathcal{M}_{\mathcal{X}}) \\
& =I(\mathbf{A},  \mathbf{B} ;\mathcal{M}_{\mathcal{X}}) + I(\mathbf{A},  \mathbf{B} ; \widetilde{A}^{\mathcal{X}}, \widetilde{B}^{\mathcal{X}}) = 0, \label{security}
\end{align}
where \eqref{security} is due to \eqref{noise1}, \eqref{noise2} and the facts that each share is encoded with $(X,S)$ Reed-Solomon code with uniformly and independently distributed noise.

Consider the privacy property, 
\small
\begin{align}
&\qquad I(Y_1,Y_2,\cdots,Y_S ;\mathbf{A},  \mathbf{B} \mid \mathbf{AB}) = I\left(\left(\mathbf{D}^{l,k}_{i}\right)_{l\in[\ell],k\in[K_c],i\in[R']}, (J_{x})_{x\in[R'K_c+2X-1]};\mathbf{A},  \mathbf{B} \mid \mathbf{AB}\right) \label{p0}\\
&=I\left(\left(\mathbf{D}^{l,k}_{i}\right)_{l\in[\ell],k\in[K_c],i\in[R']};\mathbf{A},  \mathbf{B} \mid \mathbf{AB}\right) + I\left((J_{x})_{x\in[R'K_c+2X-1]};\mathbf{A},  \mathbf{B} \mid \mathbf{AB}, \left(\mathbf{D}^{l,k}_{i}\right)_{l\in[\ell],k\in[K_c],i\in[R']} \right) \\
& = I\left((J_{x})_{x\in[R'K_c+2X-1]};\mathbf{A},  \mathbf{B} \mid \mathbf{AB}, \left(\mathbf{D}^{l,k}_{i}\right)_{l\in[\ell],k\in[K_c],i\in[R']} \right) \label{p1}\\
&\le  I\left((J_{x})_{x\in[R'K_c+2X-1]};\mathbf{A},  \mathbf{B},  \mathbf{AB}, \left(\mathbf{D}^{l,k}_{i}\right)_{l\in[\ell],k\in[K_c],i\in[R']} \right)  \\
&\le I\left(\mathcal{Z}^{server}_1, \mathcal{Z}^A, \mathcal{Z}^B;\mathbf{A},  \mathbf{B}, \left(\mathbf{D}^{l,k}_{i}\right)_{l\in[\ell],k\in[K_c],i\in[R']} \right) = 0, 
\end{align}
\normalsize
where \eqref{p0} holds because the map from $\left(\left(\mathbf{D}^{l,k}_{i}\right)_{l\in[\ell],k\in[K_c],i\in[R']}, (J_{x})_{x\in[R'K_c+2X-1]}\right)$ to $(Y_1, \cdots, Y_S)$ is bijective. Equation \eqref{p1} holds due to \eqref{noise2} and the fact $\left(\mathbf{C}^{l,k}_{i}\right)_{l\in[\ell], k\in[K_c], i\in\mathcal{E}}$ are functions of $\mathbf{AB}$. 

Consider the communication cost. The source upload cost $U_A=\frac{S}{K_cpm}$ and $U_B=\frac{S}{K_cpn}$. The server communication cost $CC=\frac{S-1}{\ell K_cmn}$.  Note that the master is able to recover $Lmn$ desired symbols from $R$ downloaded symbols, the master download cost is $D=\frac{R}{Lmn}=\frac{pmn(\ell+1)K_c+2X-1}{\ell K_cmn}$. Thus the desired costs are achievable. 

Now let us consider the computation complexity. Note that the source encoding procedure can be regarded as products of confluent Cauchy matrices by vectors. So by fast algorithms \cite{Olshevsky_Shokrollahi}, the encoding complexity of $(\mathcal{C}_{eA},\mathcal{C}_{eB})=\left(\widetilde{\mathcal{O}}\left(\frac{\lambda\kappa S\log^2S}{K_cpm}\right),  \widetilde{\mathcal{O}}\left(\frac{\kappa\mu S\log^2S}{K_cpn}\right)\right)$ is achievable. For the  server computation complexity, each server multiplies the $\ell$ pairs of shares $\widetilde{A}^s_{l}, \widetilde{B}^s_{l}, l\in[\ell]$, and returns the sum of these $\ell$ products and structured noise $ \widetilde{M}_s$. With straightforward matrix multiplication algorithms, each of the $\ell$ matrix products has a computation complexity of $\mathcal{O}\left(\frac{\lambda\kappa\mu}{pmn}\right)$ for a total of $\mathcal{O}\left(\frac{\ell\lambda\kappa\mu}{pmn}\right)$. The complexity of summation over the products and noise is $\mathcal{O}\left(\frac{\ell\lambda\mu}{mn}\right)$. To construct the noise, one server needs to encode the noise, whose complexity is $\widetilde{\mathcal{O}}\left(\frac{\lambda\mu S\log^2 S}{mn}\right)$ by fast algorithms \cite{Olshevsky_Shokrollahi}. Normalized by the number of servers, it is $\widetilde{\mathcal{O}}\left(\frac{\lambda\mu \log^2 S}{mn}\right)$. Considering these $3$ procedures, upon normalization by $L=\ell K_c$, it yields a complexity of 
$\mathcal{O}\left(\frac{\lambda\kappa\mu}{K_cpmn}\right)+\mathcal{O}\left(\frac{\lambda\mu}{K_cmn}\right)+\widetilde{\mathcal{O}}\left(\frac{\lambda\mu\log^2 S}{\ell K_c mn}\right)$ per server. The master decoding complexity is inherited from that of GCSA codes \cite{Jia_Jafar_CDBC}, which is at most $\widetilde{\mathcal{O}}(\lambda\mu p \log^2R)$. This completes the proof of Theorem \ref{thm:gcsa}.

{\it Remark: } When $L=\ell=K_c=1$, $S=R$, by setting $f_{1,1}=0$, our construction of shares of $\widetilde{A}^s$ and $\widetilde{B}^s$ essentially recovers the construction of shares in \cite{Nodehi_Maddah_MPC}.

\section{Discussion and Conclusion} \label{s:dis}
In this paper, the class of GCSA codes is expanded by including noise-alignment, so that the resulting GCSA-NA code is a solution for secure coded multi-party computation of massive matrix multiplication. For two sources and matrix multiplication, GCSA-NA strictly generalizes PS \cite{Nodehi_Maddah_MPC} and outperforms it in several key aspects. This construction also settles the asymptotic capacity of symmetric $X$-secure $T$-private information retrieval. The idea of noise-alignment can be applied to construct a scheme for $N$ sources based on $N$-CSA codes, and be combined with Strassen's construction. As open problems, exploring the optimal amount of randomness and finding the communication efficient schemes for arbitrary polynomial are interesting directions.

Since Strassen's algorithm \cite{Strassen} is an important fast matrix multiplication approach, it is interesting to show noise alignment can be combined with it for secure multi-party matrix multiplication. Consider an example with two $ 2\times 2$ block matrices $\mathbf{A}, \mathbf{B}$ and $X=1$. It can be shown that the general recursive Strassen's algorithm also works similarly. The desired product $\mathbf{C} = \left[\begin{matrix}\mathbf{C}_{1,1} ~ \mathbf{C}_{1,2}\\ \mathbf{C}_{2,1} ~ \mathbf{C}_{2,2}\end{matrix}\right]$. The Strassen's constuction constructs 14 matrices $P_i, Q_i, i\in [7]$ ($P_i$ only depends on $\mathbf{A}$ and $Q_i$ only depends on $\mathbf{B}$) and
\small
\begin{align}
    \left[\begin{matrix} \mathbf{C}_{1,1} \\
                        \mathbf{C}_{1,2} \\
                        \mathbf{C}_{2,1} \\
                        \mathbf{C}_{2,2}
        \end{matrix}\right] = \left[\begin{matrix}
        0 & -1 & 0 & 1 & 1 & 1 & 0 \\
        1 & 1 & 0 & 0 & 0 & 0 & 0 \\
        0 & 0 & 1 & 1 & 0 & 0 & 0 \\
        1 & 0 & -1 & 0 & 1 & 0 & -1
        \end{matrix}\right] \left[\begin{matrix} P_1Q_1 \\
                        P_2Q_2 \\
                        \vdots \\
                        P_7Q_7. \label{matrix0}
        \end{matrix}\right] 
\end{align}
\normalsize
This is the basic Strassen algorithm. Now let us see how we apply CSA and noise alignment to it. Each share is constructed based on CSA code principles with noise, i.e.,
\small
\begin{align}
\widetilde{A}=\Delta\left(\sum_{i \in [7]}\frac{P_i}{f_i-\alpha}+\mathbf{Z}^A\right), ~~\widetilde{B}=\sum_{i \in [7]}\frac{Q_i}{f_i-\alpha}+\mathbf{Z}^B, ~~\widetilde{A}\widetilde{B}=\sum_{i \in [7]}\frac{c_i}{f_i - \alpha}P_iQ_i + \sum^{7}_{i=0} \alpha ^i I_i. \label{exr}
\end{align}
\normalsize

If the servers directly return $\widetilde{A}\widetilde{B}$ to the master, additional information about the input may be leaked due to interference terms $P_1Q_1, \cdots, P_7Q_7$ and $\sum^{6}_{i=0} \alpha ^i I_i$. We secure the scheme by the addition of noise. The idea is that we want the master to decode $T_1,\cdots, T_7$ instead of $P_1Q_1, \cdots, P_7Q_7$, such that 
\begin{align}
    H(\mathbf{C} \mid T_1, \cdots, T_7) = 0, ~~ I(\mathbf{A}, \mathbf{B} ; T_1, \cdots, T_7 \mid \mathbf{C}) = 0.
\end{align}
$T_{1}, \cdots, T_{v}$ are constructed as follows.
\small
\begin{align}
    T_1 = P_1Q_1-\mathbf{Z}_1-\mathbf{Z}_2+\mathbf{Z}_3, ~~T_2 = P_2Q_2-\mathbf{Z}_1+\mathbf{Z}_2-\mathbf{Z}_3, ~~ T_3 = P_3Q_3-\mathbf{Z}_1,\\
    T_4 = P_4Q_4+\mathbf{Z}_1, ~~ T_5 = P_5Q_5+\mathbf{Z}_2,~~ T_6 = P_6Q_6-\mathbf{Z}_3, ~~ T_7 = P_7Q_7+\mathbf{Z}_3,
\end{align}
\normalsize
where $\mathbf{Z}_1, \mathbf{Z}_{2}, \mathbf{Z}_{3}$ are i.i.d. uniform noise matrices. To align the noise, we construct $\widetilde{Z}$,
\small
\begin{align}
    \widetilde{Z}= &\left(-\frac{c_1}{f_1-\alpha}- \frac{c_2}{f_2-\alpha} - \frac{c_3}{f_3-\alpha} + \frac{c_4}{f_4-\alpha}\right)\mathbf{Z}_1 \notag + \left(-\frac{c_1}{f_1-\alpha} + \frac{c_2}{f_2-\alpha} + \frac{c_5}{f_5-\alpha}\right)\mathbf{Z}_2 \notag \\
    &+\left(\frac{c_1}{f_1-\alpha}-\frac{c_2}{f_2-\alpha} - \frac{c_6}{f_6-\alpha}+\frac{c_7}{f_7-\alpha}\right)\mathbf{Z}_3 + \sum_{i=0}^6 \alpha^i \mathbf{Z}_{i+4},
\end{align}
\normalsize
where $\mathbf{Z}_4, \cdots, \mathbf{Z}_{10}$ are i.i.d. uniform noise matrices. The answer returned by each server to the master is $\widetilde{A}\widetilde{B}+\widetilde{Z}$. The correctness and privacy are easily proved. 

\section{Acknowledgement} This work is supported in part by funding from NSF grants CNS-1731384 and CCF-1907053, ONR grant N00014-18-1-2057 and ARO grant W911NF-19-1-0344. 




\IEEEtriggeratref{8}


\IEEEtriggeratref{52}

\bibliographystyle{IEEEtran}
\bibliography{Thesis}

\begin{thebibliography}{10}
\providecommand{\url}[1]{#1}
\csname url@samestyle\endcsname
\providecommand{\newblock}{\relax}
\providecommand{\bibinfo}[2]{#2}
\providecommand{\BIBentrySTDinterwordspacing}{\spaceskip=0pt\relax}
\providecommand{\BIBentryALTinterwordstretchfactor}{4}
\providecommand{\BIBentryALTinterwordspacing}{\spaceskip=\fontdimen2\font plus
\BIBentryALTinterwordstretchfactor\fontdimen3\font minus
  \fontdimen4\font\relax}
\providecommand{\BIBforeignlanguage}[2]{{%
\expandafter\ifx\csname l@#1\endcsname\relax
\typeout{** WARNING: IEEEtran.bst: No hyphenation pattern has been}%
\typeout{** loaded for the language `#1'. Using the pattern for}%
\typeout{** the default language instead.}%
\else
\language=\csname l@#1\endcsname
\fi
#2}}
\providecommand{\BIBdecl}{\relax}
\BIBdecl

\bibitem{Lee_Lam_Pedarsani}
K.~Lee, M.~Lam, R.~Pedarsani, D.~Papailiopoulos, and K.~Ramchandran, ``Speeding
  up distributed machine learning using codes,'' \emph{IEEE Transactions on
  Information Theory}, vol.~64, no.~3, pp. 1514--1529, 2017.

\bibitem{Yu_Maddah-Ali_Avestimehr_Polynomial}
Q.~Yu, M.~A. Maddah-Ali, and A.~S. Avestimehr, ``{Polynomial Codes: an Optimal
  Design for High-Dimensional Coded Matrix Multiplication},'' \emph{arXiv
  preprint arXiv:1705.10464}, 2017.

\bibitem{Dutta_Fahim_Haddadpour}
S.~Dutta, M.~Fahim, F.~Haddadpour, H.~Jeong, V.~Cadambe, and P.~Grover, ``{On
  the Optimal Recovery Threshold of Coded Matrix Multiplication},'' \emph{IEEE
  Transactions on Information Theory}, vol.~66, no.~1, pp. 278--301, 2020.

\bibitem{GPolyDot}
S.~Dutta, Z.~Bai, H.~Jeong, T.~Low, and P.~Grover, ``{A Unified Coded Deep
  Neural Network Training Strategy Based on Generalized PolyDot Codes for
  Matrix Multiplication},'' \emph{ArXiv:1811.10751}, Nov. 2018.

\bibitem{Yu_Maddah-Ali_Avestimehr}
Q.~Yu, M.~A. Maddah-Ali, and A.~S. Avestimehr, ``{Straggler Mitigation in
  Distributed Matrix Multiplication: Fundamental Limits and Optimal Coding},''
  \emph{IEEE Transactions on Information Theory}, vol.~66, no.~3, pp.
  1920--1933, 2020.

\bibitem{Yu_Lagrange}
Q.~Yu, S.~Li, N.~Raviv, S.~M.~M. Kalan, M.~Soltanolkotabi, and S.~Avestimehr,
  ``{Lagrange Coded Computing: Optimal Design for Resiliency, Security and
  Privacy},'' \emph{ArXiv:1806.00939}, 2018.

\bibitem{Yu_cubic}
Q.~Yu and A.~S. Avestimehr, ``{Entangled Polynomial Codes for Secure, Private,
  and Batch Distributed Matrix Multiplication: Breaking the ``Cubic
  Barrier''},'' \emph{ArXiv:2001.05101}, 2020.

\bibitem{Reisizadeh_Prakash_Pedarsani}
A.~Reisizadeh, S.~Prakash, R.~Pedarsani, and A.~S. Avestimehr, ``{Coded
  Computation over Heterogeneous Clusters},'' \emph{IEEE Transactions on
  Information Theory}, vol.~65, no.~7, pp. 4227--4242, 2019.

\bibitem{Lee_Suh_Ramchandran}
K.~Lee, C.~Suh, and K.~Ramchandran, ``High-dimensional coded matrix
  multiplication,'' in \emph{2017 IEEE International Symposium on Information
  Theory (ISIT)}.\hskip 1em plus 0.5em minus 0.4em\relax IEEE, 2017, pp.
  2418--2422.

\bibitem{Dutta_Cadambe_Short}
S.~Dutta, V.~Cadambe, and P.~Grover, ``Short-dot: Computing large linear
  transforms distributedly using coded short dot products,'' in \emph{Advances
  In Neural Information Processing Systems}, 2016, pp. 2100--2108.

\bibitem{Dutta_Cadambe_Codedconv}
------, ``Coded convolution for parallel and distributed computing within a
  deadline,'' \emph{arXiv preprint arXiv:1705.03875}, 2017.

\bibitem{Yu_Maddah-Ali_CodedDFT}
Q.~Yu, M.~A. Maddah-Ali, and A.~S. Avestimehr, ``Coded fourier transform,''
  \emph{arXiv preprint arXiv:1710.06471}, 2017.

\bibitem{Jahani-Nezhad_Maddah-Ali}
T.~Jahani-Nezhad and M.~A. Maddah-Ali, ``Codedsketch: A coding scheme for
  distributed computation of approximated matrix multiplications,'' \emph{arXiv
  preprint arXiv:1812.10460}, 2018.

\bibitem{Baharav_Lee_Ocal}
T.~Baharav, K.~Lee, O.~Ocal, and K.~Ramchandran, ``Straggler-proofing
  massive-scale distributed matrix multiplication with d-dimensional product
  codes,'' in \emph{2018 IEEE International Symposium on Information Theory
  (ISIT)}.\hskip 1em plus 0.5em minus 0.4em\relax IEEE, 2018, pp. 1993--1997.

\bibitem{Suh_Lee_Msparse}
G.~Suh, K.~Lee, and C.~Suh, ``Matrix sparsification for coded matrix
  multiplication,'' in \emph{2017 55th Annual Allerton Conference on
  Communication, Control, and Computing (Allerton)}.\hskip 1em plus 0.5em minus
  0.4em\relax IEEE, 2017, pp. 1271--1278.

\bibitem{Wang_Liu_CLT}
S.~Wang, J.~Liu, N.~Shroff, and P.~Yang, ``Fundamental limits of coded linear
  transform,'' \emph{arXiv preprint arXiv:1804.09791}, 2018.

\bibitem{Mallick_Chaudhari_Joshi}
A.~Mallick, M.~Chaudhari, U.~Sheth, G.~Palanikumar, and G.~Joshi, ``Rateless
  codes for near-perfect load balancing in distributed matrix-vector
  multiplication,'' \emph{Proc. ACM Meas. Anal. Comput. Syst.}, vol.~3, no.~3,
  2019.

\bibitem{Wang_Liu_Sparse}
S.~Wang, J.~Liu, and N.~Shroff, ``Coded sparse matrix multiplication,''
  \emph{arXiv preprint arXiv:1802.03430}, 2018.

\bibitem{Severinson_iAmat_Rosnes}
A.~Severinson, A.~G. i~Amat, and E.~Rosnes, ``Block-diagonal and lt codes for
  distributed computing with straggling servers,'' \emph{IEEE Transactions on
  Communications}, vol.~67, no.~3, pp. 1739--1753, 2018.

\bibitem{Haddadpour_Cadambe_Finite}
F.~Haddadpour and V.~R. Cadambe, ``Codes for distributed finite alphabet
  matrix-vector multiplication,'' in \emph{2018 IEEE International Symposium on
  Information Theory (ISIT)}.\hskip 1em plus 0.5em minus 0.4em\relax IEEE,
  2018, pp. 1625--1629.

\bibitem{Sheth_Dutta_Chaudhari}
U.~Sheth, S.~Dutta, M.~Chaudhari, H.~Jeong, Y.~Yang, J.~Kohonen, T.~Roos, and
  P.~Grover, ``An application of storage-optimal matdot codes for coded matrix
  multiplication: Fast k-nearest neighbors estimation,'' in \emph{2018 IEEE
  International Conference on Big Data (Big Data)}.\hskip 1em plus 0.5em minus
  0.4em\relax IEEE, 2018, pp. 1113--1120.

\bibitem{Jeong_Ye_Grover}
H.~Jeong, F.~Ye, and P.~Grover, ``Locally recoverable coded matrix
  multiplication,'' in \emph{2018 56th Annual Allerton Conference on
  Communication, Control, and Computing (Allerton)}.\hskip 1em plus 0.5em minus
  0.4em\relax IEEE, 2018, pp. 715--722.

\bibitem{Kim_Sohn_Moon_Group}
M.~Kim, J.-y. Sohn, and J.~Moon, ``Coded matrix multiplication on a group-based
  model,'' \emph{arXiv preprint arXiv:1901.05162}, 2019.

\bibitem{Park_Lee_Sohn}
H.~Park, K.~Lee, J.-y. Sohn, C.~Suh, and J.~Moon, ``Hierarchical coding for
  distributed computing,'' \emph{arXiv preprint arXiv:1801.04686}, 2018.

\bibitem{Li_Maddah-Ali_Fog}
S.~Li, M.~A. Maddah-Ali, and A.~S. Avestimehr, ``Coding for distributed fog
  computing,'' \emph{IEEE Communications Magazine}, vol.~55, no.~4, pp. 34--40,
  2017.

\bibitem{Chang_Tandon}
W.~Chang and R.~Tandon, ``On the capacity of secure distributed matrix
  multiplication,'' \emph{IEEE Global Communications Conference (GLOBECOM)},
  pp. 1--6, 2018.

\bibitem{Kakar_Ebadifar_Sezgin_CSA}
J.~Kakar, S.~Ebadifar, and A.~Sezgin, ``On the capacity and
  straggler-robustness of distributed secure matrix multiplication,''
  \emph{IEEE Access}, vol.~7, pp. 45\,783--45\,799, 2019.

\bibitem{Oliveira_Rouayheb_Karpuk}
R.~G. D’Oliveira, S.~E. Rouayheb, and D.~Karpuk, ``Gasp codes for secure
  distributed matrix multiplication,'' \emph{IEEE Transactions on Information
  Theory}, 2020, early access, DOI: 10.1109/TIT.2020.2975021.

\bibitem{Kim_Lee}
M.~Kim and J.~Lee, ``Private secure coded computation,'' \emph{IEEE
  Communications Letters}, vol.~23, no.~11, pp. 1918--1921, 2019.

\bibitem{Aliasgari_Simeone_Kliewer}
M.~Aliasgari, O.~Simeone, and J.~Kliewer, ``Distributed and private coded
  matrix computation with flexible communication load,'' \emph{arXiv preprint
  arXiv:1901.07705}, 2019.

\bibitem{Sun_Jafar_PIR}
H.~Sun and S.~A. Jafar, ``{The Capacity of Private Information Retrieval},''
  \emph{IEEE Transactions on Information Theory}, vol.~63, no.~7, pp.
  4075--4088, July 2017.

\bibitem{Sun_Jafar_TPIR}
------, ``{The Capacity of Robust Private Information Retrieval with Colluding
  Databases},'' \emph{IEEE Transactions on Information Theory}, vol.~64, no.~4,
  pp. 2361--2370, April 2018.

\bibitem{Banawan_Ulukus}
K.~Banawan and S.~Ulukus, ``{The Capacity of Private Information Retrieval from
  Coded Databases},'' \emph{IEEE Transactions on Information Theory}, vol.~64,
  no.~3, pp. 1945--1956, 2018.

\bibitem{Banawan_Ulukus_BPIR}
K.~{Banawan} and S.~{Ulukus}, ``The capacity of private information retrieval
  from byzantine and colluding databases,'' \emph{IEEE Transactions on
  Information Theory}, vol.~65, no.~2, pp. 1206--1219, Feb 2019.

\bibitem{Kadhe_Garcia_Heidarzadeh_Rouayheb_Sprintson}
S.~Kadhe, B.~Garcia, A.~Heidarzadeh, S.~E. Rouayheb, and A.~Sprintson,
  ``Private information retrieval with side information,'' \emph{IEEE
  Transactions on Information Theory}, vol.~66, no.~4, pp. 2032--2043, 2020.

\bibitem{Wang_Skoglund_SPIRAd}
Q.~Wang and M.~Skoglund, ``Secure symmetric private information retrieval from
  colluding databases with adversaries,'' \emph{2017 55th Annual Allerton
  Conference on Communication, Control, and Computing (Allerton)}, pp.
  1083--1090, 2017.

\bibitem{Jia_Sun_Jafar_XSTPIR}
Z.~{Jia}, H.~{Sun}, and S.~A. {Jafar}, ``{Cross Subspace Alignment and the
  Asymptotic Capacity of $X$ -Secure $T$ -Private Information Retrieval},''
  \emph{IEEE Trans. on Info. Theory}, vol.~65, no.~9, pp. 5783--5798, Sep.
  2019.

\bibitem{Jia_Jafar_GXSTPIR}
Z.~Jia and S.~A. Jafar, ``{On the Asymptotic Capacity of $ X $-Secure $ T
  $-Private Information Retrieval with Graph Based Replicated Storage},''
  \emph{ArXiv:1904.05906}, 2019.

\bibitem{Jia_Jafar_MDSXSTPIR}
------, ``{$X$-secure $T$-private Information Retrieval from MDS Coded Storage
  with Byzantine and Unresponsive Servers},'' \emph{ArXiv:1908.10854}, 2019.

\bibitem{Jia_Jafar_CDBC}
Z.~Jia and S.~Jafar, ``Cross-subspace alignment codes for coded distributed
  batch computation,'' \emph{ArXiv:1909.13873}, 2019.

\bibitem{Nodehi_Maddah_MPC}
H.~A. Nodehi and M.~A. Maddah-Ali, ``Secure coded multi-party computation for
  massive matrix operations,'' \emph{ArXiv:1908.04255}, 2019.

\bibitem{Yao_first}
A.~C. Yao, ``Protocols for secure computations (extended abstract),''
  \emph{23rd Annual Symposium on Foundations of Computer Science}, pp.
  160--164, 1982.

\bibitem{IA_NA}
W.~Zhao, X.~Ming, S.~Mikael, and P.~H.~Vincent, ``Secure degrees of freedom of
  wireless x networks using artificial noise alignment,'' \emph{IEEE
  Transactions on communications}, vol.~63, no.~7, pp. 2632--2646, 2015.

\bibitem{Strassen}
V.~Strassen, ``{Gaussian elimination is not optimal},'' \emph{Numerische
  Mathematik}, vol.~13, no.~4, pp. 354--356, 1969.

\bibitem{Nodehi_MPC_EP_ISIT}
H.~A. Nodehi and M.~A. Maddah-Ali, ``Limited-sharing multi-party computation
  for massive matrix operations,'' \emph{IEEE International Symposium on
  Information Theory}, 2018.

\bibitem{Nodehi_MPC_EP}
H.~A. Nodehi, S.~R.~H. Najarkolaei, and M.~A. Maddah-Ali, ``Entangled
  polynomial coding in limited-sharing multi-party computation,'' \emph{IEEE
  Information Theory Workshop}, 2018.

\bibitem{Gasca_Martinez_Muhlbach}
M.~Gasca, J.~Martinez, and G.~M{\"u}hlbach, ``{Computation of Rational
  Interpolants with Prescribed Poles},'' \emph{Journal of Computational and
  Applied Mathematics}, vol.~26, no.~3, pp. 297--309, 1989.

\bibitem{Olshevsky_Shokrollahi}
V.~Olshevsky and A.~Shokrollahi, ``{A Superfast Algorithm for Confluent
  Rational Tangential Interpolation Problem via Matrix-Vector Multiplication
  for Confluent Cauchy-like Matrices},'' \emph{Contemporary Mathematics}, vol.
  280, pp. 31--46, 2001.

\end{thebibliography}

\end{document}